\documentclass[twocolumn]{aa}

\usepackage{graphicx}
\usepackage{hyperref}
\usepackage{threeparttablex} 
\usepackage{lscape} 
\usepackage{url}
\usepackage{tablefootnote}
\usepackage{makecell}
\usepackage[dvipsnames]{xcolor}

\usepackage{txfonts}
\begin{document}

\title{A compendium of distances to molecular clouds in the Star Formation Handbook \thanks{Table A.1 is available in electronic form at the CDS via anonymous ftp to \url{cdsarc.u-strasbg.fr (130.79.128.5)} or via \url{http://cdsarc.u-strasbg.fr/viz-bin/qcat?J/A+A/}. It is also available on the Harvard Dataverse at \url{https://doi.org/10.7910/DVN/07L7YZ}}}

\authorrunning{Zucker et al.}
\titlerunning{A Compendium of Distances to Molecular Clouds in the Star Formation Handbook}
   
\author{Catherine Zucker\inst{1},
    Joshua S. Speagle\inst{1},
    Edward F. Schlafly\inst{2},
    Gregory M. Green\inst{3},
    Douglas P. Finkbeiner\inst{1},
    Alyssa Goodman\inst{1,5},
    Jo\~{a}o Alves\inst{4,5}}
          
\institute{Center for Astrophysics $\vert$ Harvard \& Smithsonian, 60 Garden St., Cambridge, MA 02138, USA \label{inst1}\\
    \email{catherine.zucker@cfa.harvard.edu, jspeagle@cfa.harvard.edu}
    \and Lawrence Berkeley National Laboratory, One Cyclotron Road, Berkeley, CA 94720, USA\label{inst2} 
    \and Kavli Institute for Particle Astrophysics and Cosmology, Physics and Astrophysics Building, 452 Lomita Mall, Stanford, CA 94305, USA \label{inst3}
    \and University of Vienna, Department of Astrophysics, T{\"u}rkenschanzstra{\ss}e 17, 1180 Vienna, Austria \label{inst4}
    \and Radcliffe Institute for Advanced Study, Harvard University, 10 Garden St, Cambridge, MA 02138\label{inst5}}
 
\abstract
{Accurate distances to local molecular clouds are critical for understanding the star and planet formation process, yet distance measurements are often obtained inhomogeneously on a cloud-by-cloud basis. We have recently developed a method which combines stellar photometric data with \textit{Gaia} DR2 parallax measurements in a Bayesian framework to infer the distances of nearby dust clouds to a typical accuracy of $\sim5\%$. After refining the technique to target lower latitudes and incorporating deep optical data from DECam in the southern Galactic plane, we have derived a catalog of distances to molecular clouds in Reipurth (2008, Star Formation Handbook, vols I and II) which contains a large fraction of the molecular material in the solar neighborhood. Comparison with distances derived from maser parallax measurements towards the same clouds shows our method produces consistent distances with $\lesssim10\%$ scatter for clouds across our entire distance spectrum (150 pc $-$ 2.5 kpc). We hope this catalog of homogeneous distances will serve as a baseline for future work.}


\maketitle

\section{Introduction}

The Star Formation Handbook, divided into two volumes for the Northern \citep{Northern_Handbook} and Southern \citep{Southern_Handbook} sky, contains around sixty of the most important star forming regions within 2 kpc. Written by a team of 105 authors, the Handbook spans 1900 pages, and includes the most comprehensive discussion of individual low- and high-mass star forming regions published to date. Since the proximity of these clouds facilitates high-resolution observations across the electromagnetic spectrum, together these regions inform much of our knowledge of how molecular gas is transformed into stars. 

Characterizing the specifics of this process relies on robust distance estimates to star-forming regions, and while many of the Handbook's clouds are well studied, their distances are not well constrained. Several clouds in the Handbook have distance estimates in the literature that vary by at least a factor of two (e.g. Circinus Molecular Cloud, North America Nebula, Coalsack Nebula, NGC 2362, IC 5146), while many others (e.g. Lagoon Nebula, Pipe Nebula, IC2944, NGC2264) may show better agreement, but with large distance uncertainties ($\gtrsim 30\%$).

In this work, we leverage the technique presented in \citet{Zucker_2019} to produce a supplementary catalog of distances to molecular clouds in the Star Formation Handbook \citep{Northern_Handbook, Southern_Handbook} with a typical distance uncertainty of $\approx 5\%$. Our method relies on the colors of stars, taking advantage of the fact that stars behind a dust screen appear redder. An alternative method of determining the presence of dust is to track stellar number counts, rather than colors, as dust clouds obscure some fraction of background stars. This latter method was pioneered by Max Wolf in the early twentieth century, when he established a technique for estimating the distances to dark nebulae using the apparent magnitudes of stars \citep{Wolf_1923}. In what is now known as a ``Wolf diagram", Wolf plotted the number of stars per unit solid angle versus their apparent magnitudes in both extinguished and unextinguished regions towards the nebulae. Under the assumption that all stars have the same absolute magnitude, Wolf determined the distance to dark nebulae by characterizing the apparent magnitude at which one observes a drop in stellar density towards the obscured sightlines \citep[see e.g. discussion in Chapter 6 of][]{Shore_2002}.

A more precise study of cloud distances based on stellar photometry requires modeling the colors of stars and their types, and in a more modern sense, our methodology is similar to that presented by \citet{Neckel_1980}, which has a rich history in the literature \citep[see e.g.][]{Schlafly_2014, Sale_2018, Lallement_2019, Green_2019, Marshall_2006, Rezaei_2018, Yan_2019}. By combining \textit{Gaia} DR2 parallax measurements with stellar photometry, we infer the distance, extinction, type, and $R_V$ of stellar sources in sightlines towards local molecular clouds. Unlike \citet{Wolf_1923}, we require stars to be detected both in front of and behind the cloud, and we fit a simple line-of-sight dust model to the set of \textit{Gaia}-constrained stellar distance and extinction estimates to infer the distance at which we observe a ``break" in stellar reddening. 

While \citet{Zucker_2019} provided a uniform catalog of distances to over twenty-five named clouds, it did not incorporate deep optical data in the southern Galactic plane (e.g. towards the Southern Coalsack, Circinus, IC 2944), nor was the technique intended to target clouds near $b=0^\circ$, particularly towards the inner galaxy (e.g. M16, M17, M20, NGC6604). Here, we refine the technique to target approximately thirty additional named regions selected from the Star Formation Handbook. When combined with the results of \citet{Zucker_2019}, this includes distance estimates to almost every major cloud in \citet{Northern_Handbook, Southern_Handbook}. In Sect. \ref{methods}, we briefly summarize the methodology presented in \citet{Zucker_2019} to infer distances to each cloud along with updates to data and methods implemented in this work. In Sect. \ref{catalog}, we present our new catalog of distances to clouds in the Star Formation Handbook, as well as an interactive 3-D figure of the entire distance catalog. A machine readable version of the catalog is available on the Harvard Dataverse\footnote{\textcolor{blue}{\href{https://doi.org/10.7910/DVN/07L7YZ}{https://doi.org/10.7910/DVN/07L7YZ}}} and will be made available via the CDS. In Sect. \ref{discussion}, we compare our dust-based distances from stars to gas-based distances from masers, finding good agreement between the two independent methods. Finally, we conclude in Sect. \ref{conclusion}. 


\section{Data and methods} \label{methods}

Our technique is identical to \citet{Zucker_2019} save for three improvements summarized in Sect. \ref{improvements}. Here, we recapitulate the core data products and methodology we employ from \citet{Zucker_2019} to infer the per-star distance extinction measurements and the line-of-sight dust distribution. For a detailed description of the data and methods, see Sect. 2 and Sect. 3 in \citet{Zucker_2019}. 

In brief, we derive the distance, extinction, type, and $R_V$ towards stars along sightlines through local molecular clouds using optical and near-infrared photometry. For sightlines above a declination $\delta=-30^\circ$, we use optical point spread function (PSF) photometry from the PanSTARRS1 survey \citep{Magnier_2016, Chambers_2016} and PSF near-infrared photometry from the 2MASS survey \citep{Skrutskie_2006}. For stars below a $\delta=-30^\circ$ and outside the southern Galactic plane (Galactic latitude $\vert b \vert > 5^\circ$), we use optical aperture photometry from the NOAO Source Catalog \citep[NSC;][]{Nidever_18} and near-infrared data from the 2MASS survey. Finally, for stars in the Southern galactic plane ($\delta< -30^\circ$ and $\vert b \vert  < 5^\circ$) we employ optical PSF photometry from the DECam Galactic Plane Survey \citep[DECaPS;][]{Schlafly_2018} and near-infrared data from the 2MASS survey. When available, we cross-match sources with \textit{Gaia} \citep{Brown_2018} to obtain parallax measurements for each star. A \textit{Gaia} parallax measurements is not required for inclusion, but is incorporated as an additional Gaussian likelihood term. The incorporation of \textit{Gaia} has the largest effect for stars with high signal-to-noise parallax measurements in the solar neighborhood, but is also able to resolve distance degeneracies for stars with low signal-to-noise parallax measurements beyond a few kiloparsecs \citep[for more details, see Sect. 3.1 and interactive Figure 1 of][]{Zucker_2019}. Stars are selected in either $0.7^\circ$ beams (for nearby clouds) or $0.2^\circ$ beams (for more distant clouds) centered on sightlines of interest through each cloud, as discussed further in Sect. \ref{sample}.

\subsection{Per-star inference} \label{perstar}

We model the observed magnitudes $\mathbf{\hat{m}}$ of the stars\footnote{The stellar templates we employ are derived using photometry in the native magnitude system of each survey. The PS1, DECaPS, and NSC surveys use AB magnitudes, while 2MASS uses Vega magnitudes. As a result, the magnitude system of the observed photometry $\mathbf{\hat{m}}$ depends on the available bands.} (in the optical and near-infrared bands) as a function of distance, extinction, stellar type, and $R_V$ using a technique similar to that outlined in \citet{Green_2014, Green_2015, Green_2018}:
\begin{equation}
\mathbf{\hat{m}} = \mathbf{m}_{\rm int}(M_r, {\rm [Fe/H]}) + A_V \times (\mathbf{R} + R_V \times \mathbf{R}') + \mu
\end{equation}
Here, $\mathbf{m}_{\rm int}$ is the set of intrinsic (un-reddened) magnitudes for the star as a function of stellar type, $A_V$ is the extinction, $R_V$ is the ``differential extinction'', $\mathbf{R}$ and $\mathbf{R}'$ characterize the overall reddening as a function of magnitude, and $\mu$ is the distance modulus. The intrinsic colors of stars are based on a set of empirical templates derived from fitting a stellar locus in a low reddening region of the sky; these templates parameterize the star's colors as a function of its metallicity ([Fe/H]) and absolute magnitude in the PanSTARRS1 r-band ($M_r$). The baseline and differential reddening ``vectors'' $\mathbf{R}$ and $\mathbf{R}'$ are derived using the results from \citet{Schlafly_2016} and are identical to those used in \citet{Zucker_2019}.

For northern clouds ($\delta > -30^\circ$), the stellar templates and reddening curve we use are identical to those employed in \citet{Green_2019}. For the southern clouds, we transform these templates into the DECaPS $grizy$ bands using color transformations derived on low-reddening calibration fields. For the NSC data, we additionally apply zero-point corrections derived on similar calibration fields to bring their measurements in line with the AB photometric system. The reddening vectors are converted to the DECam system by integrating the interpolated curve from \citet{Schlafly_2016} through the relevant DECam filter set. See Appendix \ref{details} for more details.

The posterior probability, $P(\theta | \hat{\mathbf{m}}, \hat{\varpi})$, that our observed magnitudes $\hat{\mathbf{m}}$ are consistent with the predicted model magnitudes $\mathbf{m}(\theta) \equiv \mathbf{m}(M_r, {\rm [Fe/H]}, A_V, R_V, \mu)$ and the measured \textit{Gaia} parallax measurement $\hat{\varpi}$ is based on Bayes' Theorem:
\begin{equation}
P(\theta | \hat{\mathbf{m}}, \hat{\varpi}) \propto P(\hat{\mathbf{m}} | \theta) \, P(\hat{\varpi} | \mu) \, P(\theta)
\end{equation}
The probability of observing our given magnitudes $P(\hat{\mathbf{m}} | \theta)$ and corresponding parallax $P(\hat{\varpi} | \mu)$ is based on Gaussian measurement noise. The prior probability $\pi(\theta)$ of our underlying parameters $\theta$ incorporates previous work on the luminosity, number density, and metallicity of stars in the Galaxy \citep{Bressan_2012, Ivezic_2008, Juric_2008} as well as the variation of the optical-infrared extinction curve across the Milky Way disk \citep{Schlafly_2016}.

Like \citet{Zucker_2019}, we fit for our model parameters ($M_r, {\rm [Fe/H]}, A_V, R_V, \mu$) using the public code \texttt{brutus}\footnote{The brutus source code is available on \href{https://github.com/joshspeagle/brutus/}{GitHub} as well as Zenodo: \url{https://doi.org/10.5281/zenodo.3348370
}} (Speagle et al., in prep). We marginalize over stellar type ($M_r$, [Fe/H]) and $R_V$ to obtain the 2-D distance-extinction posterior $P(\mu, A_V | \hat{\mathbf{m}}, \hat{\varpi})$ for each star, which is subsequently used in our line-of-sight fit.

\begin{figure}
    \begin{center}
        \includegraphics[width=\hsize]{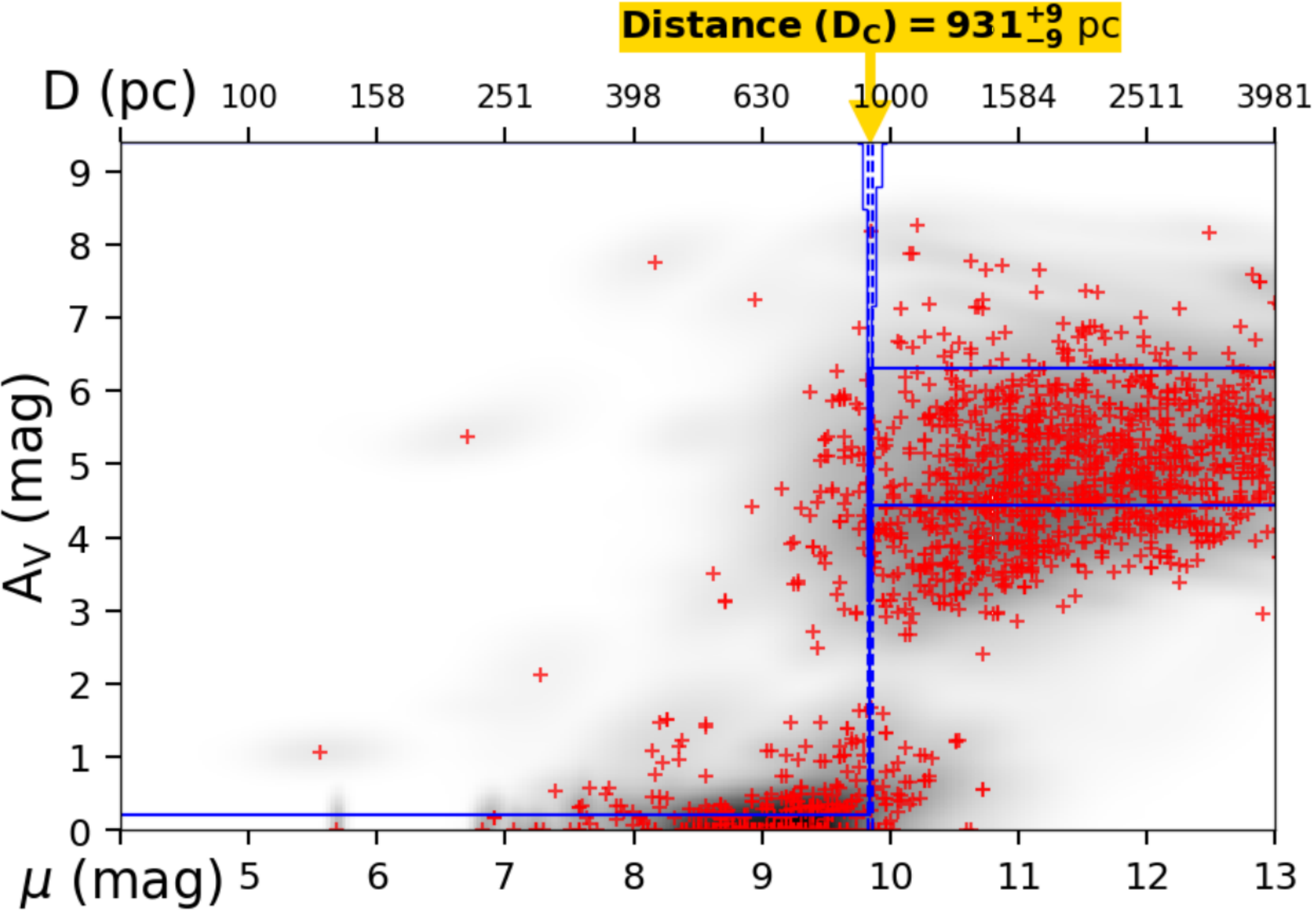}
        \caption{Our line-of-sight extinction model as a function of distance towards the Vela C cloud ($l,b = 264.7^\circ, 1.4^\circ$). The background grayscale shows the probabilistic distance and extinction estimates for all the stars used in the fit, with the most probable distance and extinction to every star marked via a red cross. The blue step function shows the typical extinction profile inferred for the sightline. The median foreground extinction prior to the cloud distance $D_C$ (at distance modulus $\mu_C$) is shown as a blue horizontal line. The range of estimated distances to the cloud is shown as an inverted blue histogram, with the median cloud distance and the corresponding 16th/84th percentiles marked with solid and dashed blue vertical lines, respectively. Beyond the cloud distance, the bottom and top blue lines show the 16th and 84th percentile of the \textit{Planck}-based extinction towards the stars. The distance uncertainties do not include systematic uncertainties, which we estimate to be roughly $5\%$.}
        \label{fig:sample_profile}
    \end{center}
\end{figure}

\subsection{Line-of-sight inference}

We fit a simple line-of-sight dust model to the set of stellar distance-extinction posteriors to determine the distance to each cloud. We model each cloud as a thin dust screen at a particular distance modulus $\mu_C$. The total extinction through the screen towards each star $i$ is based on two components. The first is the extinction due to the dust cloud parameterized by $N \times C_i$, where $C_i$ is the reddening towards the star at 353 GHz based on measurements from \textit{Planck} \citep{Planck_2014} times the star's typical $R_V$ (see Sect. \ref{perstar}) and $N$ is a normalization constant. The second is due to a small amount of foreground extinction $f$ caused by nearby dust unassociated with the cloud. The line-of-sight extinction as a function of distance towards a star in each sightline of interest is then:
\begin{equation}
\label{eq:profile}
A_V(\mu) = \left\{
        \begin{array}{lr}
            f &  \mu <  \mu_C \\
            f + N \times C^i &  \mu \geq \mu_C \\
        \end{array}
    \right.
\end{equation}

In addition to the simplistic model above, we also allow for additional scatter in the extinction from star to star in the foreground ($s_{\rm fore}$) and background ($s_{\rm back}$) of each sightline. Finally, we reject outlying stars unassociated with the cloud based on an adaptive threshold ($P_b$) as part of a Gaussian mixture model. 

The probability of our cloud having a particular set of parameters $\alpha=\{\mu_C, N, f, s_{\rm fore}, s_{\rm back}, P_b \}$ is based on a prior $P(\alpha)$ as well as the distance and extinction of each of the $n$ stars in the cloud. To account for measurement uncertainties from each star, we marginalize over all possible distances and extinctions to get:
\begin{equation}
\label{eq:like}
P(\alpha|\{\hat{\mathbf{m}}_i\}_{i=1}^{n}) \propto P(\alpha) \prod_{i=1}^{n} \int P(\mu_i,A_V^i|\alpha) \, P(\mu_i, A_V^i|\hat{\mathbf{m}}_i) \, d\mu_i dA_v^i
\end{equation}
where $P(\mu_i, A_V^i | \hat{\mathbf{m}}_i)$ is the posterior for each star based on Sect.\ref{perstar} and $P(\mu_i, A_V^i|\alpha)$ is the probability of observing a given distance and extinction given our line-of-sight model. Our prior $P(\alpha)$ follows the same functional forms as those used in \citet{Zucker_2019} except for two small changes. First, we have increased our allowed range for the normalization factor $N$ (relative to \textit{Planck}) to be from $0.01$ to $2.0$ to better account for larger amounts of background dust. Second, we have increased the upper bound of $f$ to be 50\% of the mean stellar-based extinction to the cloud to account for greater amounts of foreground material. We find these two changes allows us to substantially better model behavior in the Galactic plane at $b\sim0^\circ$. We caution that for sightlines with low normalization factors, the spatial information from \textit{Planck} is not providing any additional constraint on the distance. \textit{Planck} clearly provides a poor description of the data in these cases, and a flat dust template would work equally well in these scenarios. Nevertheless, the jump in extinction is large enough that the templates have little effect on the distance determination, and are simply employed to remain consistent with other, less extinguished sightlines where the spatial information is expected to play a larger role. 

We sample for our line-of-sight model parameters $\alpha$ using the public nested sampling code \texttt{dynesty} \citep{Speagle_2019} with the same setup as described in \citet{Zucker_2019} \footnote{See their Appendix C}. 

\subsection{Sample selection} \label{sample}

About 25 of the 60 clouds highlighted in the Star Formation Handbook are targeted in \citet{Zucker_2019}. This constitutes $\approx 125$ sightlines towards some of the most famous nearby clouds (e.g. Perseus, Taurus, Orion, CMa OB1, California, etc.). Here we target $\approx 30$ new clouds using 200 new sightlines. Following \citet{Zucker_2019} and guided by discussion of each region in the Star Formation Handbook, we choose representative sightlines in and around the clouds of interest that are well suited to the technique described above. Our technique requires actually seeing stars both in front of and behind each cloud, so we consequently avoid targeting particularly dense regions ($E(B-V) \gtrsim 5$ mag) in favor of lower density envelopes ($0.15 \; {\rm mag} < E(B-V) < 5 \; {\rm mag}$). Consequently, while these sightlines are generally near the traditional star-forming cores and clumps associated with each star-forming region, they are not centered on them.

We apply three general cuts to improve the quality of our stellar sample. First, we restrict our sample to stars lying in pixels with $E(B-V) > 0.15$ mag (based on \textit{Planck} estimates at 353 GHz). Second, we remove stars whose photometry is inconsistent with our stellar modeling at the $2-3 \sigma$ level (equivalent to a p-value < 0.01). Finally, novel to this work, we apply a cut to remove stars whose inferred distance is inconsistent with its \textit{Gaia} parallax measurement (when available) at the $2\sigma$ level. This occurs primarily when our modeling mis-identifies faraway giant stars towards the Galactic plane (which have measured parallaxes $\hat{\varpi} \sim 0$) as nearby dwarfs.

As in \citet{Zucker_2019}, we adopt two different samples of stars to determine distances in different regimes. For faraway clouds ($d \gtrsim 300$ pc), we select all stars within a $0.2^\circ$ beam. For nearby clouds ($d\approx 100-300$ pc), we instead select only M-dwarf stars using a set of color and magnitude cuts \citep[the same employed in][]{Zucker_2019} within a larger $0.7^\circ$ beam. For these very nearby clouds, this M-dwarf cut prevents the large number of background stars from overwhelming the sparse number of foreground stars.\footnote{The exceptions are the southern clouds Circinus and Norma, which lie at $\approx 700$ pc; the ``near" technique was used for these clouds because they lie in the Galactic plane, and the higher stellar density combined with the greater depth of the DECaPS survey meant the number of background stars quickly overwhelmed the number of foreground stars with the ``faraway" technique, which led to an unreliable fit. }

\subsection{Summary of updates to \citet{Zucker_2019}} \label{improvements}

To summarize, our methodology is largely similar to \citet{Zucker_2019}, save three improvements:

\begin{itemize}
\item We extend our technique to the southern Galactic plane by incorporating deep optical data from the DECam Galactic Plane Survey \citep[DECaPS;][]{Schlafly_2018}. The first DECaPS data release targeted the sky at $\delta< -30^\circ$ and $\vert b \vert <  5^\circ $. Reaching typical single-exposure depths of 23.7, 22.8, 22.3, 21.9, and 21.0 mag in the $g$, $r$, $i$, $z$, and $Y$ bands, respectively, the survey obtained PSF photometry for around 2 billion stars, allowing us to expand our distance catalog to southern clouds including the Southern Coalsack, Carina, IC2944, Vela C, RCW38, Ara, Circinus, and Norma.

\item We modify the priors on our line-of-sight fit to accommodate greater amounts of unassociated foreground dust and background dust along the line-of-sight to improve our modeling in the Galactic plane.

\item We identify and remove giant stars misclassified as dwarfs by our stellar modeling using a parallax-based cut.
\end{itemize}

\noindent The full catalog of distances is presented in Sect. \ref{catalog}. 


\section{Catalog} \label{catalog}

\begin{figure*}[t]
    \begin{center}
        \includegraphics[width=0.95\hsize]{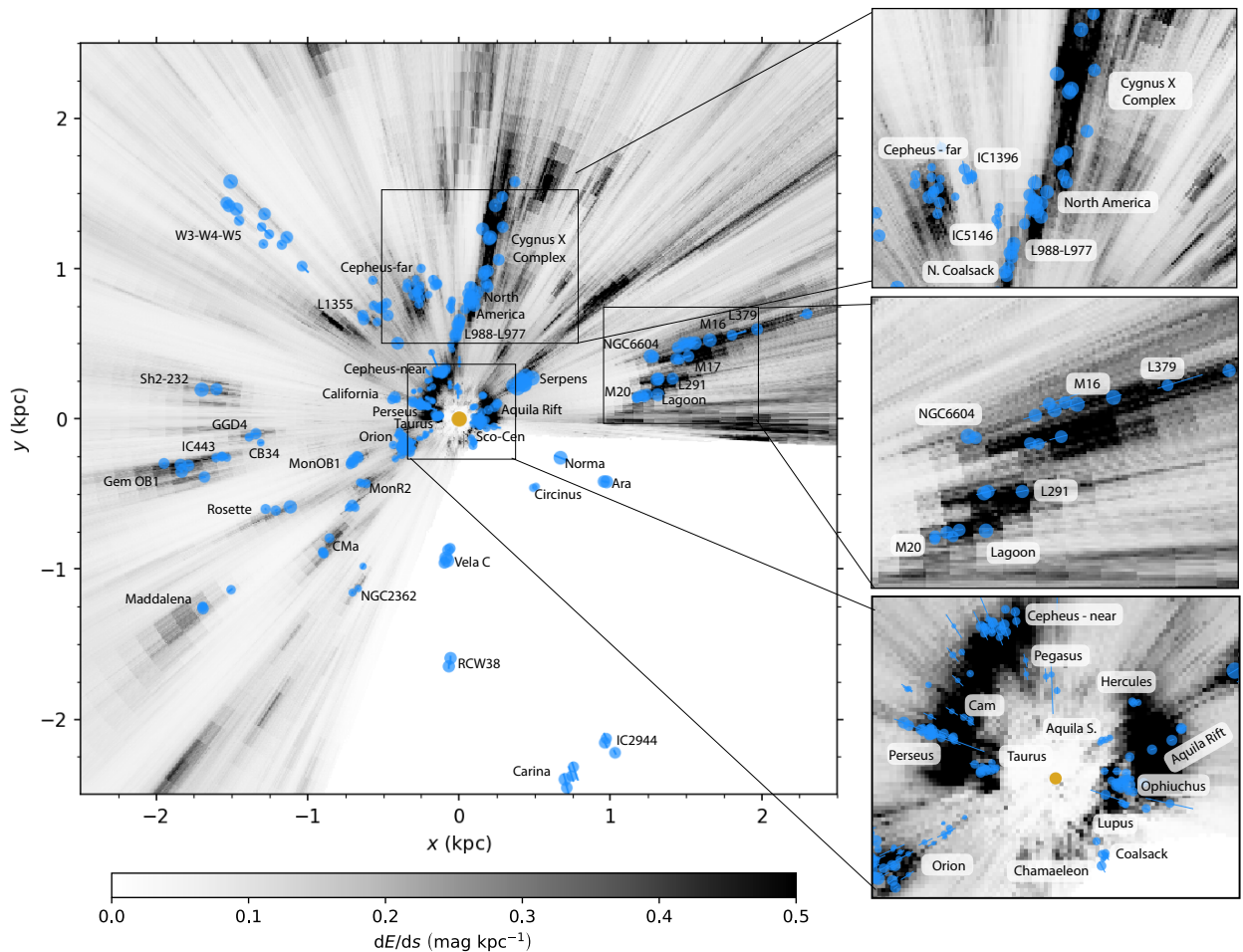}
        \caption{A bird's-eye view of the Star Formation Handbook cloud catalog (colored blue points), looking down on the Galactic disk with the sun (orange circle) at the center. The catalog is overlaid on the 3D ``Bayestar19'' dust map from \citet{Green_2019} integrated from $z= \pm 300$ pc off the plane. The points have been arbitrarily scaled according to their dust extinction (between $A_V=0$ mag and $A_V = 9$ mag), so larger scatter points indicate more extinguished sightlines. The statistical errors (corresponding to the 16th/84th percentile of the cloud distances) are indicated via the line segments; an additional systematic uncertainty is expected, as reported in Table \ref{tab:distance}. The right-hand panels show zoom-ins of the clouds towards $l=90^\circ$ (top), clouds near the Sagittarius arm (middle), and clouds within the nearest 375 pc of the sun (bottom). In case this 3D figure does not render, an interactive 3D version is also accessible online at \href{https://faun.rc.fas.harvard.edu/czucker/Paper_Figures/handbook_distances.html}{\color{blue}{https://faun.rc.fas.harvard.edu/czucker/Paper\_Figures/handbook\_distances.html}}.}
        \label{topdown}
    \end{center}
\end{figure*}

In Table \ref{tab:distance}, we summarize our distance results for the 326 sightlines targeted towards clouds in the Star Formation Handbook. We include the central coordinates of the sightline, the named region the sightline is associated with, and the volume (Northern or Southern) and page number for the Star Formation Handbook chapter in which it is discussed. A machine-readable version of this table is available for download on the Dataverse and at the CDS \footnote{\textcolor{blue}{\href{https://doi.org/10.7910/DVN/07L7YZ}{https://doi.org/10.7910/DVN/07L7YZ}}}, and also includes values for the ancillary model parameters determined for each sightline. On the Dataverse, we also include a set of complementary figures for each sightline, including 1-D and 2-D marginalized posterior distributions of our model parameters and how the line-of-sight extinction varies as a function of distance towards each cloud \footnote{\textcolor{blue}{\href{https://doi.org/10.7910/DVN/SBFNG7}{https://doi.org/10.7910/DVN/SBFNG7}}}. An example figure, showing how extinction varies as a function of distance for the Vela C cloud (towards $l, b = 264.7^\circ, 1.4^\circ$) is shown in Figure \ref{fig:sample_profile}.

The distances we report in Table \ref{tab:distance} represent the median cloud distance for each sightline determined using the set of distance samples returned by \texttt{dynesty} \citep{Speagle_2019}. The first set of error bars represent our $1\sigma$ statistical errors, computed using the 16th and 84th percentiles of our distance samples. The statistical uncertainties are usually low (on the order of 1-2\%), with the exception of sightlines with very few foreground stars, where the fit is poorly constrained, resulting in higher statistical uncertainties. The second set of errorbars is our systematic uncertainty, estimated to be 5\% in distance (see discussion in \citet{Zucker_2019}), though we caution here, as in \citet{Zucker_2019}, that this systematic uncertainty is likely higher for clouds at farther distances (beyond $\approx$ 1.5 kpc) or with more complicated line-of-sight dust structures, due to the simplicity of our line-of-sight dust model. We also anticipate higher systematic uncertainties for clouds in the NSC footprint, as discussed in Sect. \ref{details}. These systematic uncertainties are reflected in Table \ref{tab:distance}. We recommend the statistical and systematic uncertainties be added in quadrature. 

Our results cover most chapters in the Star Formation Handbook. The primary exceptions are open clusters or OB associations which are not associated with significant and extended dust emission, thereby rendering them unsuitable for this technique. These include:
\begin{itemize}
    \item the young open clusters NGC 6383 (Southern Volume, p. 497), NGC 6231 (Southern Volume, p. 401), and $\sigma$-Orionis (Northern Volume, p. 732),
    \item the extended OB association Sco OB2 (Southern Volume, p. 235),
    \item the chapter on Young Nearby Loose Associations (Southern Volume, p. 757), and
    \item the chapter on Dispersed Young Population in Orion (Northern Volume, p. 838) 
\end{itemize}
Certain clusters, like the LkH$\alpha$ 101 cluster (part of NGC 1579), are not specifically targeted, but are known from the literature to be associated with the larger California Molecular Cloud \citep{Lada_2009}, to which we are able to obtain a reliable distance. We likewise exclude the chapter Star Formation in Bok Globules and Small Clouds (Southern Volume, p. 847) as the size of many globules is equivalent to the pixel scale of our dust templates from \textit{Planck}; a tailored approach employing higher angular resolution templates (e.g. from \textit{Herschel}) would be better suited for these objects, provided enough background stars could be seen through the globules. While we targeted NGC 6334 (Southern Volume, p. 456), RCW 120 (Southern Volume, p. 437), and GM 24 (Southern Volume, p. 449), these regions were simply too extinguished (near $b=0^\circ$) and too confused (towards the Galactic center and beyond 1 kpc) that we were unable to obtain a reliable distance to them. 

Finally, we note that a few clouds in Table \ref{tab:distance} are not explicitly mentioned in the Handbook (e.g. Pegasus, Hercules, Aquila South, Spider, Draco, Maddalena). These were targeted in \citet{Zucker_2019} and are included here for completeness.


\section{Discussion} \label{discussion}

\begin{figure}
    \begin{center}
        \includegraphics[width=0.88\hsize]{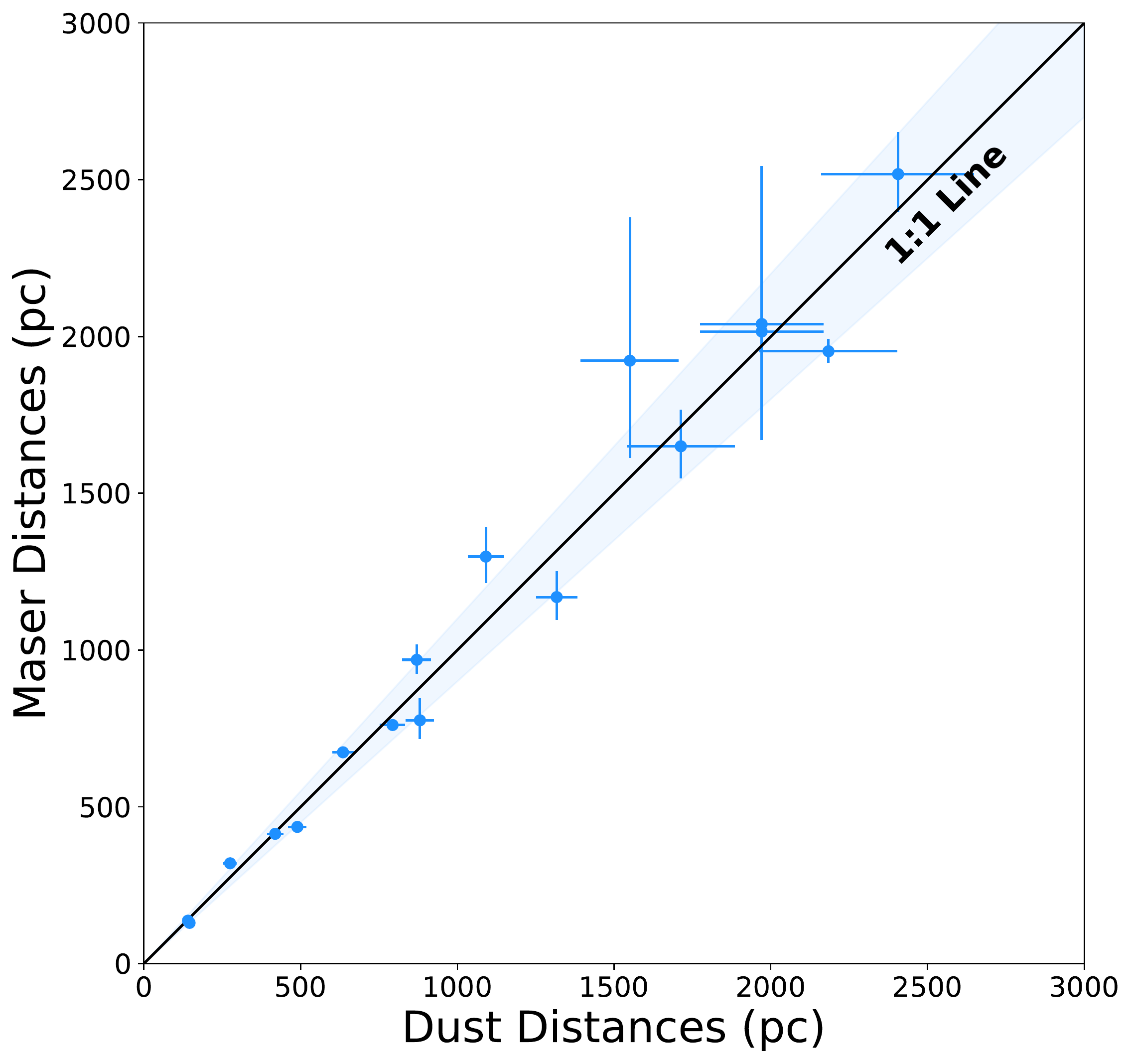}
        \caption{Comparison of our average dust distances derived from stars to VLBI distances (mainly masers) derived from trigonometric parallax observations towards low- and high-mass star forming regions. We find good agreement between the two methods, with a typical scatter of $\lesssim 10\%$ in distance for clouds across the entire distance range explored (100 pc $-$ 2.5 kpc). The errors we report for the dust distances account for both the statistical and systematic uncertainty, added in quadrature. The shaded region bounding the 1:1 indicates the typical combined uncertainty we estimate for our dust distances}
        \label{masercomp}
    \end{center}
\end{figure}

In Figure \ref{topdown} we show a bird's-eye view of our distance catalog, overlaid on the 3-D ``Bayestar19'' dust map from \citet{Green_2019} integrated over $z = \pm 300$ pc from the plane. As expected, our dust distances are broadly consistent with the \citet{Green_2019} 3D dust map, with any small discrepancies arising from the fact that our technique is optimized for molecular cloud distance determination, while the \citet{Green_2019} technique characterizes all the dust along the line of sight. As is apparent in Figure \ref{topdown}, our distance catalog includes a majority of the molecular emission (and associated dust) out to 2.5 kpc. Much of the catalog is not randomly distributed, but instead forms coherent quasi-linear complexes such as those seen in the nearby Sco-Cen clouds consisting of Serpens, Aquila, Lupus, Chamaeleon, and Corona Australis. However, we do see some evidence of the ``fingers of God" effect, owing to our angular resolution being much finer in comparison to our distance resolution. This effect is clearly seen towards the W3, W4, and W5 star-forming regions, where we obtain similar average distances to recent results obtained using \textit{Gaia} DR2 data for OB stars in the complex \citep{Navarete_2019} but with significantly more scatter along the line-of-sight. In the caption of Figure \ref{topdown} we also link to an interactive 3-D version of this figure, where users can pan, zoom, and hover over any sightline to view the name of the cloud.

To gauge the accuracy of our catalog, we compare our star- and dust-based distances with those obtained independently using trigonometric parallax observations of masers in low-mass and high-mass star-forming regions. Based on the BeSSeL \citep{Reid_2014, Reid_2016, Brunthaler_2011} and GOBELINS surveys \citep{Loinard_2013}, we identify sightlines within a projected distance of 50 pc from VLBI source positions in the following regions\footnote{Since maser emission is rare and highly variable in low-mass star forming regions, the GOBELINS survey is not preferentially targeting masers, and we include any published VLBI observation of compact radio emission from the GOBELINS survey associated with YSOs \citep[e.g.][]{Ortiz_Leon_2018a}}:
\begin{itemize}
    \item Taurus \citep{Galli_2018},
    \item Ophiuchus \citep{Ortiz_Leon_2017a}, 
    \item Perseus \citep{Ortiz_Leon_2018a}, 
    \item Serpens/Aquila \citep{Ortiz_Leon_2017b}, 
    \item Gemini OB1 \citep[``IRAS06061", ``S252";][]{Reid_2014}, 
    \item Monoceros R2 \citep[``G213.70-12.6";][]{Reid_2016}, 
    \item Sh2-232 \citep[``G173.48+2.44"][]{Sakai_2019}, 
    \item L379 \citep[``G016.86-2.15"][]{Reid_2016}\footnote{This distance will appear in Rygl et al. (in prep.), and is publicly available on the BeSSeL website as an accompanying data product from
\citet{Reid_2016}, which forms the core of the maser catalog comparison. See \url{http://bessel.vlbi-astrometry.org/bayesian}}, 
    \item L977 and L988 \citep[``G090.21+2.32";][]{Xu_2013}, 
    \item NGC2362 \citep[``VYCMa"][]{Zhang_2012}, 
    \item S106 \citep[``G076.38-0.61"][]{Xu_2013}, 
    \item Orion \citep{Reid_2014}, 
    \item Cepheus-Far \citep[``IRAS22198", ``G108.18+5.51"][]{Reid_2014}, and 
    \item the W3 star-forming region \citep[``W3OH";][]{Reid_2014}.
\end{itemize}
While we determine distances to the same regions, each approach targets different parts of the clouds using different observations: the maser distances are obtained using radio data towards the most extinguished sightlines while our star- and dust-based distances are obtained using optical to near-infrared data towards the lower density envelopes. 

One small complication is that our method can often obtain multiple distance estimates across a single cloud, while maser parallax measurements are usually limited to one or two per star forming region. To facilitate a fair comparison, we calculate the average dust distance for every maser above using only sightlines within a projected distance of 50 pc from the maser sources on the plane-of-the-sky, based on each cloud's maser distance. 

The relationship between our average dust distances and the maser distances is shown in Figure \ref{masercomp}. Overall, we find good agreement between the two methods. Across the range of distances explored, the typical scatter between our average dust distance compared to the respective maser distance is just under 10\%. While this is consistent with our estimated uncertainties for faraway clouds, this is a few percent higher than our estimated uncertainties for nearby clouds. This could indicate that we slightly underestimate our uncertainties (by a few percent) or that the masers are not capturing cloud substructure that is present in our averaged dust distances, leading to discrepancies in the cloud distances caused by, for example, distance gradients in the clouds themselves. 

Nevertheless, we find no systematic difference in the distances derived from each method. This lack of systematic distance offset is evident in Figure \ref{masercomp}, with the data providing a very good fit to the 1:1 line. The good agreement underlines the accuracy of this method with respect to the traditional standard of cloud distance determination. While we are not able to target the most extinguished sightlines, our technique is relatively inexpensive, does not require a radio source, and can be applied over a much larger fraction of each star-forming region. This heralds future opportunities to study the precise 3D dust structure of these clouds in finer detail, which is currently not possible with maser parallax observations.



\section{Conclusion} \label{conclusion}

Using the technique presented in \citet{Zucker_2019}, we obtain accurate distances to $\approx 60$ star-forming regions within 2.5 kpc described in the Star Formation Handbook \citep{Northern_Handbook, Southern_Handbook}. Averaged over a molecular cloud, we find that our dust distances agree with traditional maser-based distances to within $\approx 10\%$ with no discernable systematic offsets. Our catalog contains famous molecular cloud associations (e.g. the Sco-Cen clouds) as well as other possible structures that will be the study of future work. A machine-readable version of the full catalog is publicly available on the Dataverse and at the CDS \footnote{\textcolor{blue}{\href{https://doi.org/10.7910/DVN/07L7YZ}{https://doi.org/10.7910/DVN/07L7YZ}}}. Upcoming data releases from \textit{Gaia}, in combination with future all-sky deep optical surveys \citep[e.g. LSST;][]{LSST_2009}, should present exciting new opportunities to further improve these distances, in pursuit of better 3D maps of molecular clouds in the solar neighborhood.


\begin{acknowledgements}

We would like to thank our referee, John Bally, and the Star Formation Handbook editor, Bo Reipurth for their feedback, the implementation of which greatly improved the quality of this manuscript. \\

This work took part under the program Milky-Way-\textit{Gaia} of the PSI2 project funded by the IDEX Paris-Saclay, ANR-11-IDEX-0003-02.\\

The computations in this paper utilize resources from the Odyssey cluster, which is supported by the FAS Division of Science, Research Computing Group at Harvard University. \\

The visualization, exploration, and interpretation of data presented in this work was made possible using the glue visualization software, supported under NSF grant OAC-1739657. \\

The interactive component of Figure 2 was created using the visualization software plot.ly (\url{plot.ly}). \\

D.P.F. and C.Z. acknowledge support by NSF grant AST-1614941, ``Exploring the Galaxy: 3-Dimensional Structure and Stellar Streams.'' \\

C.Z. and J.S.S. are supported by the NSF Graduate Research Fellowship Program (Grant No. 1650114) and the Harvard Data Science Initiative. \\

E.S. acknowledges support for this work by NASA through ADAP grant NNH17AE75I and Hubble Fellowship grant HST-HF2-51367.001-A awarded by the Space Telescope Science Institute, which is operated by the Association of Universities for Research in Astronomy, Inc., for NASA, under contract NAS 5-26555. \\

DECaPS is based on observations at Cerro Tololo Inter-American Observatory, National Optical Astronomy Observatory (NOAO Prop. ID: 2014A-0429, 2016A0327, 2016B-0279, 2018A-0251, 2018B-0271, and 2019A-0265; PI: Finkbeiner), which is operated by the Association of Universities for Research in Astronomy (AURA) under a cooperative agreement with the National Science Foundation. \\

The Pan-STARRS1 Surveys (PS1) and the PS1 public science archive have been made possible through contributions by the Institute for Astronomy, the University of Hawaii, the Pan-STARRS Project Office, the Max-Planck Society and its participating institutes, the Max Planck Institute for Astronomy, Heidelberg and the Max Planck Institute for Extraterrestrial Physics, Garching, The Johns Hopkins University, Durham University, the University of Edinburgh, the Queen's University Belfast, the Harvard-Smithsonian Center for Astrophysics, the Las Cumbres Observatory Global Telescope Network Incorporated, the National Central University of Taiwan, the Space Telescope Science Institute, the National Aeronautics and Space Administration under Grant No. NNX08AR22G issued through the Planetary Science Division of the NASA Science Mission Directorate, the National Science Foundation Grant No. AST-1238877, the University of Maryland, Eotvos Lorand University (ELTE), the Los Alamos National Laboratory, and the Gordon and Betty Moore Foundation. \\

This publication makes use of data products from the Two Micron All Sky Survey, which is a joint project of the University of Massachusetts and the Infrared Processing and Analysis Center/California Institute of Technology, funded by the National Aeronautics and Space Administration and the National Science Foundation. \\

NOAO is operated by the Association of Universities for Research in Astronomy (AURA) under a cooperative agreement with the National Science Foundation. Database access and other data services are provided by the NOAO Data Lab. \\

This project used data obtained with the Dark Energy Camera (DECam), which was constructed by the Dark Energy Survey (DES) collaboration. Funding for the DES Projects has been provided by the U.S. Department of Energy, the U.S. National Science Foundation, the Ministry of Science and Education of Spain, the Science and Technology Facilities Council of the United Kingdom, the Higher Education Funding Council for England, the National Center for Supercomputing Applications at the University of Illinois at Urbana-Champaign, the Kavli Institute of Cosmological Physics at the University of Chicago, Center for Cosmology and Astro-Particle Physics at the Ohio State University, the Mitchell Institute for Fundamental Physics and Astronomy at Texas A\&M University, Financiadora de Estudos e Projetos, Funda{\c c}\~ao Carlos Chagas Filho de Amparo, Financiadora de Estudos e Projetos, Funda{\c c}\~ao Carlos Chagas Filho de Amparo \`a Pesquisa do Estado do Rio de Janeiro, Conselho Nacional de Desenvolvimento Cient\'ifico e Tecnol\'ogico and the Minist\'erio da Ci\^encia, Tecnologia e Inova{\c c}\~ao, the Deutsche Forschungsgemeinschaft and the Collaborating Institutions in the Dark Energy Survey. The Collaborating Institutions are Argonne National Laboratory, the University of California at Santa Cruz, the University of Cambridge, Centro de Investigaciones En\'ergeticas, Medioambientales y Tecnol\'ogicas–Madrid, the University of Chicago, University College London, the DES-Brazil Consortium, the University of Edinburgh, the Eidgen\"ossische Technische Hochschule (ETH) Z\"urich, Fermi National Accelerator Laboratory, the University of Illinois at Urbana-Champaign, the Institut de Ci\`encies de l'Espai (IEEC/CSIC), the Institut de F\'isica d'Altes Energies, Lawrence Berkeley National Laboratory, the Ludwig-Maximilians Universit\"at M\"unchen and the associated Excellence Cluster Universe, the University of Michigan, the National Optical Astronomy Observatory, the University of Nottingham, the Ohio State University, the University of Pennsylvania, the University of Portsmouth, SLAC National Accelerator Laboratory, Stanford University, the University of Sussex, and Texas A\&M University. \\

This work has made use of data from the European Space Agency (ESA) mission {\it Gaia} (\url{https://www.cosmos.esa.int/gaia}), processed by the {\it Gaia} Data Processing and Analysis Consortium (DPAC, \url{https://www.cosmos.esa.int/web/gaia/dpac/consortium}). Funding for the DPAC has been provided by national institutions, in particular the institutions participating in the {\it Gaia} Multilateral Agreement. \\ 

This work has made use of the following software: \texttt{astropy} \citep{Astropy_2018}, \texttt{dustmaps} \citep{Dustmaps_2018}, \texttt{bokeh} \citep{Bokeh_2018}, \texttt{plotly} \citep{plotly}, \texttt{healpy} \citep{Healpy_2005}, \texttt{dynesty} \citep{Speagle_2019}, and \texttt{glue} \citep{glueviz_2017}. \\

JSS is grateful for Rebecca Bleich's patience and support. \\

\end{acknowledgements}

\appendix 

\section{Photometric Calibration and Stellar Modeling} \label{details}

To obtain stellar templates for the southern clouds ($\delta < -30^\circ$) for both the NSC \citep{Nidever_18} and DECaPS \citep{Schlafly_2018} surveys, we start by selecting a low-reddening field (a beam of radius $5^\circ$ centered on $\alpha$, $\delta$ = $7^\circ$, $0^\circ$) in the NSC footprint that also overlaps with the Pan-STARRS1 survey.

Using this field to link NSC to PS1, we derive a color transform to convert our model intrinsic PS1-2MASS colors and absolute PS1 $r$-band magnitudes from our northern stellar templates \citep[from][]{Green_2019} to the DECam system. Specifically, we follow a similar procedure to that employed in \citet{Schlafly_2018} (see their Sect. 6.2) and model the color transformations between DECam and PS1 using a quadratic polynominal as a function of $c \equiv g_\mathrm{PS1}-i_\mathrm{PS1}$ color: 

\begin{align*}
g_\mathrm{DECam} - g_\mathrm{PS1} &=  -0.00339  +0.04430c +0.01389c^2 \\
&-0.01274c^3 + 0.00199c^4 \\
r_\mathrm{DECam} - r_\mathrm{PS1} &= -0.00155 - 0.08364c + 0.07627c^2 \\
& - 0.04278c^3 + 0.00607c^4 \\
i_\mathrm{DECam} - i_\mathrm{PS1} &= +0.00948 - 0.06805c + 0.08693c^2 \\
& - 0.05854c^3 + 0.00936^4\\
z_\mathrm{DECam} - z_\mathrm{PS1} &= +0.01007 - 0.02935c - 0.00509c^2 \\
& - 0.00012c^3 - 0.00076c^4 \\
Y_\mathrm{DECam} - y_\mathrm{PS1} &= 0.01369 - 0.03150c  + 0.01449c^2 \\
& - 0.00241c^3 - 0.00035c^4
\end{align*}

The constant terms above depend on the absolute photometric calibration, and were obtained by comparing the derived color transformations to expectations from the filter systems of PS1 and DECam following the same procedure as outlined in Sect. 5.2 of \citet{Schlafly_2018}. This procedure revealed that the NSC photometry was significantly offset from the PS1 AB system, requiring offsets of
\begin{align*}
g_\mathrm{NSC} - g_\mathrm{DECam} &= -0.006 \\
r_\mathrm{NSC} - r_\mathrm{DECam} &= 0.102 \\
i_\mathrm{NSC} - i_\mathrm{DECam} &= 0.088 \\
z_\mathrm{NSC} - z_\mathrm{DECam} &= 0.090 \\
Y_\mathrm{NSC} - y_\mathrm{DECam} &= 0.047
\end{align*}
These were applied to the NSC magnitudes before performing the stellar inference. In addition to these overall zero-points, we also found strong evidence for a magnitude-dependent trend in these offsets, with shifts of $\sim 0.03$ mag between 15-20th mag. We opt not to include an additional correction for this trend for simplicity.

As the DECaPS survey has already been calibrated to the PS1 system following the same procedure, no further photometric corrections were needed for clouds in the DECaPS footprint.

Although we have attempted to bring DECaPS and NSC onto the PS1 photometric system, there might still be systematics in the derived distance estimates between the different datasets. While there is overlap between DECaPS/NSC DECam observations and PS1 towards the Galactic center, it is difficult to compare results between the three surveys due to the extreme amount of reddening, cloud confusion, and other systematics present in those sightlines. Unfortunately, there is very limited overlap in molecular clouds in uncrowded regions where NSC aperture photometry performs well, so we are not able to confirm agreement between NSC and DECaPS in this regime.

Comparisons of DECaPS and NSC-derived distances over a low-reddening footprint give consistent distances at the 5\% level. This is on par with the overall 5\% systematic uncertainty of our technique and has been added in quadrature to the systematic distance uncertainties for the NSC-derived distances to Lupus, Corona Australis, and Chamaeleon in Table \ref{tab:distance}.


\bibliographystyle{aa}
\bibliography{aanda}

\begin{thebibliography}{49}
\expandafter\ifx\csname natexlab\endcsname\relax\def\natexlab#1{#1}\fi

\bibitem[{{Astropy Collaboration} {et~al.}(2018){Astropy Collaboration},
  {Price-Whelan}, {Sip{\'{o}}cz}, {G{\"u}nther}, {Lim}, {Crawford}, {Conseil},
  {Shupe}, {Craig}, {Dencheva}, {Ginsburg}, {VanderPlas}, {Bradley},
  {P{\'e}rez-Su{\'a}rez}, {de Val- Borro}, {Aldcroft}, {Cruz}, {Robitaille},
  {Tollerud}, {Ardelean}, {Babej}, {Bach}, {Bachetti}, {Bakanov}, {Bamford},
  {Barentsen}, {Barmby}, {Baumbach}, {Berry}, {Biscani}, {Boquien}, {Bostroem},
  {Bouma}, {Brammer}, {Bray}, {Breytenbach}, {Buddelmeijer}, {Burke},
  {Calderone}, {Cano Rodr{\'\i}guez}, {Cara}, {Cardoso}, {Cheedella}, {Copin},
  {Corrales}, {Crichton}, {D'Avella}, {Deil}, {Depagne}, {Dietrich}, {Donath},
  {Droettboom}, {Earl}, {Erben}, {Fabbro}, {Ferreira}, {Finethy}, {Fox},
  {Garrison}, {Gibbons}, {Goldstein}, {Gommers}, {Greco}, {Greenfield},
  {Groener}, {Grollier}, {Hagen}, {Hirst}, {Homeier}, {Horton}, {Hosseinzadeh},
  {Hu}, {Hunkeler}, {Ivezi{\'c}}, {Jain}, {Jenness}, {Kanarek}, {Kendrew},
  {Kern}, {Kerzendorf}, {Khvalko}, {King}, {Kirkby}, {Kulkarni}, {Kumar},
  {Lee}, {Lenz}, {Littlefair}, {Ma}, {Macleod}, {Mastropietro}, {McCully},
  {Montagnac}, {Morris}, {Mueller}, {Mumford}, {Muna}, {Murphy}, {Nelson},
  {Nguyen}, {Ninan}, {N{\"o}the}, {Ogaz}, {Oh}, {Parejko}, {Parley}, {Pascual},
  {Patil}, {Patil}, {Plunkett}, {Prochaska}, {Rastogi}, {Reddy Janga},
  {Sabater}, {Sakurikar}, {Seifert}, {Sherbert}, {Sherwood-Taylor}, {Shih},
  {Sick}, {Silbiger}, {Singanamalla}, {Singer}, {Sladen}, {Sooley},
  {Sornarajah}, {Streicher}, {Teuben}, {Thomas}, {Tremblay}, {Turner},
  {Terr{\'o}n}, {van Kerkwijk}, {de la Vega}, {Watkins}, {Weaver}, {Whitmore},
  {Woillez}, {Zabalza}, \& {Astropy Contributors}}]{Astropy_2018}
{Astropy Collaboration}, {Price-Whelan}, A.~M., {Sip{\'{o}}cz}, B.~M., {et~al.}
  2018, \aj, 156, 123

\bibitem[{{Bokeh Development Team}(2018)}]{Bokeh_2018}
{Bokeh Development Team}. 2018, Bokeh: Python library for interactive
  visualization

\bibitem[{{Bressan} {et~al.}(2012){Bressan}, {Marigo}, {Girardi}, {Salasnich},
  {Dal Cero}, {Rubele}, \& {Nanni}}]{Bressan_2012}
{Bressan}, A., {Marigo}, P., {Girardi}, L., {et~al.} 2012, \mnras, 427, 127

\bibitem[{{Brunthaler} {et~al.}(2011){Brunthaler}, {Reid}, {Menten}, {Zheng},
  {Bartkiewicz}, {Choi}, {Dame}, {Hachisuka}, {Immer}, \&
  {Moellenbrock}}]{Brunthaler_2011}
{Brunthaler}, A., {Reid}, M.~J., {Menten}, K.~M., {et~al.} 2011, Astronomische
  Nachrichten, 332, 461

\bibitem[{{Chambers} {et~al.}(2016){Chambers}, {Magnier}, {Metcalfe},
  {Flewelling}, {Huber}, {Waters}, {Denneau}, {Draper}, {Farrow}, {Finkbeiner},
  {Holmberg}, {Koppenhoefer}, {Price}, {Saglia}, {Schlafly}, {Smartt},
  {Sweeney}, {Wainscoat}, {Burgett}, {Grav}, {Heasley}, {Hodapp}, {Jedicke},
  {Kaiser}, {Kudritzki}, {Luppino}, {Lupton}, {Monet}, {Morgan}, {Onaka},
  {Stubbs}, {Tonry}, {Banados}, {Bell}, {Bender}, {Bernard}, {Botticella},
  {Casertano}, {Chastel}, {Chen}, {Chen}, {Cole}, {Deacon}, {Frenk},
  {Fitzsimmons}, {Gezari}, {Goessl}, {Goggia}, {Goldman}, {Grebel}, {Hambly},
  {Hasinger}, {Heavens}, {Heckman}, {Henderson}, {Henning}, {Holman}, {Hopp},
  {Ip}, {Isani}, {Keyes}, {Koekemoer}, {Kotak}, {Long}, {Lucey}, {Liu},
  {Martin}, {McLean}, {Morganson}, {Murphy}, {Nieto-Santisteban}, {Norberg},
  {Peacock}, {Pier}, {Postman}, {Primak}, {Rae}, {Rest}, {Riess}, {Riffeser},
  {Rix}, {Roser}, {Schilbach}, {Schultz}, {Scolnic}, {Szalay}, {Seitz},
  {Shiao}, {Small}, {Smith}, {Soderblom}, {Taylor}, {Thakar}, {Thiel},
  {Thilker}, {Urata}, {Valenti}, {Walter}, {Watters}, {Werner}, {White},
  {Wood-Vasey}, \& {Wyse}}]{Chambers_2016}
{Chambers}, K.~C., {Magnier}, E.~A., {Metcalfe}, N., {et~al.} 2016, ArXiv
  e-prints [\eprint[arXiv]{1612.05560}]

\bibitem[{{Dzib} {et~al.}(2016){Dzib}, {Ortiz-Le{\'o}n}, {Loinard},
  {Mioduszewski}, {Rodr{\'\i}guez}, {Torres}, \& {Deller}}]{Dzib_2016}
{Dzib}, S.~A., {Ortiz-Le{\'o}n}, G.~N., {Loinard}, L., {et~al.} 2016, \apj,
  826, 201

\bibitem[{{Gaia Collaboration} {et~al.}(2018){Gaia Collaboration}, {Brown},
  {Vallenari}, {Prusti}, {de Bruijne}, {Babusiaux}, {Bailer-Jones}, {Biermann},
  {Evans}, {Eyer}, \& et~al.}]{Brown_2018}
{Gaia Collaboration}, {Brown}, A.~G.~A., {Vallenari}, A., {et~al.} 2018, \aap,
  616, A1

\bibitem[{{Galli} {et~al.}(2018){Galli}, {Loinard}, {Ortiz-L{\'e}on},
  {Kounkel}, {Dzib}, {Mioduszewski}, {Rodr{\'\i}guez}, {Hartmann}, {Teixeira},
  \& {Torres}}]{Galli_2018}
{Galli}, P. A.~B., {Loinard}, L., {Ortiz-L{\'e}on}, G.~N., {et~al.} 2018, \apj,
  859, 33

\bibitem[{{G{\'o}rski} {et~al.}(2005){G{\'o}rski}, {Hivon}, {Banday},
  {Wandelt}, {Hansen}, {Reinecke}, \& {Bartelmann}}]{Healpy_2005}
{G{\'o}rski}, K.~M., {Hivon}, E., {Banday}, A.~J., {et~al.} 2005, \apj, 622,
  759

\bibitem[{{Green}(2018)}]{Dustmaps_2018}
{Green}, G. 2018, The Journal of Open Source Software, 3, 695

\bibitem[{{Green} {et~al.}(2018){Green}, {Schlafly}, {Finkbeiner}, {Rix},
  {Martin}, {Burgett}, {Draper}, {Flewelling}, {Hodapp}, {Kaiser}, {Kudritzki},
  {Magnier}, {Metcalfe}, {Tonry}, {Wainscoat}, \& {Waters}}]{Green_2018}
{Green}, G.~M., {Schlafly}, E.~F., {Finkbeiner}, D., {et~al.} 2018, \mnras,
  478, 651

\bibitem[{{Green} {et~al.}(2014){Green}, {Schlafly}, {Finkbeiner}, {Juri{\'c}},
  {Rix}, {Burgett}, {Chambers}, {Draper}, {Flewelling}, {Kudritzki}, {Magnier},
  {Martin}, {Metcalfe}, {Tonry}, {Wainscoat}, \& {Waters}}]{Green_2014}
{Green}, G.~M., {Schlafly}, E.~F., {Finkbeiner}, D.~P., {et~al.} 2014, \apj,
  783, 114

\bibitem[{{Green} {et~al.}(2015){Green}, {Schlafly}, {Finkbeiner}, {Rix},
  {Martin}, {Burgett}, {Draper}, {Flewelling}, {Hodapp}, {Kaiser}, {Kudritzki},
  {Magnier}, {Metcalfe}, {Price}, {Tonry}, \& {Wainscoat}}]{Green_2015}
{Green}, G.~M., {Schlafly}, E.~F., {Finkbeiner}, D.~P., {et~al.} 2015, \apj,
  810, 25

\bibitem[{{Green} {et~al.}(2019){Green}, {Schlafly}, {Zucker}, {Speagle}, \&
  {Finkbeiner}}]{Green_2019}
{Green}, G.~M., {Schlafly}, E.~F., {Zucker}, C., {Speagle}, J.~S., \&
  {Finkbeiner}, D.~P. 2019, arXiv e-prints, arXiv:1905.02734

\bibitem[{Inc.(2015)}]{plotly}
Inc., P.~T. 2015, plotly: Collaborative data science

\bibitem[{{Ivezi{\'c}} {et~al.}(2008){Ivezi{\'c}}, {Sesar}, {Juri{\'c}},
  {Bond}, {Dalcanton}, {Rockosi}, {Yanny}, {Newberg}, {Beers}, {Allende
  Prieto}, {Wilhelm}, {Lee}, {Sivarani}, {Norris}, {Bailer-Jones}, {Re
  Fiorentin}, {Schlegel}, {Uomoto}, {Lupton}, {Knapp}, {Gunn}, {Covey}, {Allyn
  Smith}, {Miknaitis}, {Doi}, {Tanaka}, {Fukugita}, {Kent}, {Finkbeiner},
  {Munn}, {Pier}, {Quinn}, {Hawley}, {Anderson}, {Kiuchi}, {Chen}, {Bushong},
  {Sohi}, {Haggard}, {Kimball}, {Barentine}, {Brewington}, {Harvanek},
  {Kleinman}, {Krzesinski}, {Long}, {Nitta}, {Snedden}, {Lee}, {Harris},
  {Brinkmann}, {Schneider}, \& {York}}]{Ivezic_2008}
{Ivezi{\'c}}, {\v Z}., {Sesar}, B., {Juri{\'c}}, M., {et~al.} 2008, \apj, 684,
  287

\bibitem[{{Juri{\'c}} {et~al.}(2008){Juri{\'c}}, {Ivezi{\'c}}, {Brooks},
  {Lupton}, {Schlegel}, {Finkbeiner}, {Padmanabhan}, {Bond}, {Sesar},
  {Rockosi}, {Knapp}, {Gunn}, {Sumi}, {Schneider}, {Barentine}, {Brewington},
  {Brinkmann}, {Fukugita}, {Harvanek}, {Kleinman}, {Krzesinski}, {Long},
  {Neilsen}, {Nitta}, {Snedden}, \& {York}}]{Juric_2008}
{Juri{\'c}}, M., {Ivezi{\'c}}, {\v Z}., {Brooks}, A., {et~al.} 2008, \apj, 673,
  864

\bibitem[{{Lada} {et~al.}(2009){Lada}, {Lombardi}, \& {Alves}}]{Lada_2009}
{Lada}, C.~J., {Lombardi}, M., \& {Alves}, J.~F. 2009, \apj, 703, 52

\bibitem[{{Lallement} {et~al.}(2019){Lallement}, {Babusiaux}, {Vergely},
  {Katz}, {Arenou}, {Valette}, {Hottier}, \& {Capitanio}}]{Lallement_2019}
{Lallement}, R., {Babusiaux}, C., {Vergely}, J.~L., {et~al.} 2019, \aap, 625,
  A135

\bibitem[{{Loinard}(2013)}]{Loinard_2013}
{Loinard}, L. 2013, in IAU Symposium, Vol. 289, Advancing the Physics of Cosmic
  Distances, ed. R.~{de Grijs}, 36--43

\bibitem[{{LSST Science Collaboration} {et~al.}(2009){LSST Science
  Collaboration}, {Abell}, {Allison}, {Anderson}, {Andrew}, {Angel}, {Armus},
  {Arnett}, {Asztalos}, {Axelrod}, \& et~al.}]{LSST_2009}
{LSST Science Collaboration}, {Abell}, P.~A., {Allison}, J., {et~al.} 2009,
  arXiv e-prints, arXiv:0912.0201

\bibitem[{{Magnier} {et~al.}(2016){Magnier}, {Schlafly}, {Finkbeiner}, {Tonry},
  {Goldman}, {R{\"o}ser}, {Schilbach}, {Chambers}, {Flewelling}, {Huber},
  {Price}, {Sweeney}, {Waters}, {Denneau}, {Draper}, {Hodapp}, {Jedicke},
  {Kudritzki}, {Metcalfe}, {Stubbs}, \& {Wainscoast}}]{Magnier_2016}
{Magnier}, E.~A., {Schlafly}, E.~F., {Finkbeiner}, D.~P., {et~al.} 2016, ArXiv
  e-prints [\eprint[arXiv]{1612.05242}]

\bibitem[{{Marshall} {et~al.}(2006){Marshall}, {Robin}, {Reyl{\'e}},
  {Schultheis}, \& {Picaud}}]{Marshall_2006}
{Marshall}, D.~J., {Robin}, A.~C., {Reyl{\'e}}, C., {Schultheis}, M., \&
  {Picaud}, S. 2006, \aap, 453, 635

\bibitem[{{Navarete} {et~al.}(2019){Navarete}, {Galli}, \&
  {Damineli}}]{Navarete_2019}
{Navarete}, F., {Galli}, P. A.~B., \& {Damineli}, A. 2019, \mnras, 487, 2771

\bibitem[{{Neckel} \& {Klare}(1980)}]{Neckel_1980}
{Neckel}, T. \& {Klare}, G. 1980, \aaps, 42, 251

\bibitem[{{Nidever} {et~al.}(2018){Nidever}, {Dey}, {Olsen}, {Ridgway},
  {Nikutta}, {Juneau}, {Fitzpatrick}, {Scott}, \& {Valdes}}]{Nidever_18}
{Nidever}, D.~L., {Dey}, A., {Olsen}, K., {et~al.} 2018, \aj, 156, 131

\bibitem[{{Ortiz-Le{\'o}n} {et~al.}(2017{\natexlab{a}}){Ortiz-Le{\'o}n},
  {Dzib}, {Kounkel}, {Loinard}, {Mioduszewski}, {Rodr{\'\i}guez}, {Torres},
  {Pech}, {Rivera}, \& {Hartmann}}]{Ortiz_Leon_2017b}
{Ortiz-Le{\'o}n}, G.~N., {Dzib}, S.~A., {Kounkel}, M.~A., {et~al.}
  2017{\natexlab{a}}, \apj, 834, 143

\bibitem[{{Ortiz-Le{\'o}n} {et~al.}(2018){Ortiz-Le{\'o}n}, {Loinard}, {Dzib},
  {Galli}, {Kounkel}, {Mioduszewski}, {Rodr{\'\i}guez}, {Torres}, {Hartmann},
  \& {Boden}}]{Ortiz_Leon_2018a}
{Ortiz-Le{\'o}n}, G.~N., {Loinard}, L., {Dzib}, S.~A., {et~al.} 2018, \apj,
  865, 73

\bibitem[{{Ortiz-Le{\'o}n} {et~al.}(2017{\natexlab{b}}){Ortiz-Le{\'o}n},
  {Loinard}, {Kounkel}, {Dzib}, {Mioduszewski}, {Rodr{\'\i}guez}, {Torres},
  {Gonz{\'a}lez-L{\'o}pezlira}, {Pech}, \& {Rivera}}]{Ortiz_Leon_2017a}
{Ortiz-Le{\'o}n}, G.~N., {Loinard}, L., {Kounkel}, M.~A., {et~al.}
  2017{\natexlab{b}}, \apj, 834, 141

\bibitem[{{Planck Collaboration} {et~al.}(2014){Planck Collaboration},
  {Abergel}, {Ade}, {Aghanim}, {Alves}, {Aniano}, {Armitage-Caplan}, {Arnaud},
  {Ashdown}, {Atrio-Barandela}, {Aumont}, {Baccigalupi}, {Banday}, {Barreiro},
  {Bartlett}, {Battaner}, {Benabed}, {Beno{\^\i}t}, {Benoit-L{\'e}vy},
  {Bernard}, {Bersanelli}, {Bielewicz}, {Bobin}, {Bock}, {Bonaldi}, {Bond},
  {Borrill}, {Bouchet}, {Boulanger}, {Bridges}, {Bucher}, {Burigana}, {Butler},
  {Cardoso}, {Catalano}, {Chamballu}, {Chary}, {Chiang}, {Chiang},
  {Christensen}, {Church}, {Clemens}, {Clements}, {Colombi}, {Colombo},
  {Combet}, {Couchot}, {Coulais}, {Crill}, {Curto}, {Cuttaia}, {Danese},
  {Davies}, {Davis}, {de Bernardis}, {de Rosa}, {de Zotti}, {Delabrouille},
  {Delouis}, {D{\'e}sert}, {Dickinson}, {Diego}, {Dole}, {Donzelli},
  {Dor{\'e}}, {Douspis}, {Draine}, {Dupac}, {Efstathiou}, {En{\ss}lin},
  {Eriksen}, {Falgarone}, {Finelli}, {Forni}, {Frailis}, {Fraisse},
  {Franceschi}, {Galeotta}, {Ganga}, {Ghosh}, {Giard}, {Giardino},
  {Giraud-H{\'e}raud}, {Gonz{\'a}lez-Nuevo}, {G{\'o}rski}, {Gratton},
  {Gregorio}, {Grenier}, {Gruppuso}, {Guillet}, {Hansen}, {Hanson}, {Harrison},
  {Helou}, {Henrot-Versill{\'e}}, {Hern{\'a}ndez- Monteagudo}, {Herranz},
  {Hildebrandt}, {Hivon}, {Hobson}, {Holmes}, {Hornstrup}, {Hovest},
  {Huffenberger}, {Jaffe}, {Jaffe}, {Jewell}, {Joncas}, {Jones}, {Juvela},
  {Keih{\"a}nen}, {Keskitalo}, {Kisner}, {Knoche}, {Knox}, {Kunz},
  {Kurki-Suonio}, {Lagache}, {L{\"a}hteenm{\"a}ki}, {Lamarre}, {Lasenby},
  {Laureijs}, {Lawrence}, {Leonardi}, {Le{\'o}n-Tavares}, {Lesgourgues},
  {Levrier}, {Liguori}, {Lilje}, {Linden-V{\o}rnle}, {L{\'o}pez-Caniego},
  {Lubin}, {Mac{\'\i}as-P{\'e}rez}, {Maffei}, {Maino}, {Mandolesi}, {Maris},
  {Marshall}, {Martin}, {Mart{\'\i}nez-Gonz{\'a}lez}, {Masi}, {Massardi},
  {Matarrese}, {Matthai}, {Mazzotta}, {McGehee}, {Melchiorri}, {Mendes},
  {Mennella}, {Migliaccio}, {Mitra}, {Miville-Desch{\^e}nes}, {Moneti},
  {Montier}, {Morgante}, {Mortlock}, {Munshi}, {Murphy}, {Naselsky}, {Nati},
  {Natoli}, {Netterfield}, {N{\o}rgaard-Nielsen}, {Noviello}, {Novikov},
  {Novikov}, {Osborne}, {Oxborrow}, {Paci}, {Pagano}, {Pajot}, {Paladini},
  {Paoletti}, {Pasian}, {Patanchon}, {Perdereau}, {Perotto}, {Perrotta},
  {Piacentini}, {Piat}, {Pierpaoli}, {Pietrobon}, {Plaszczynski},
  {Pointecouteau}, {Polenta}, {Ponthieu}, {Popa}, {Poutanen}, {Pratt},
  {Pr{\'e}zeau}, {Prunet}, {Puget}, {Rachen}, {Reach}, {Rebolo}, {Reinecke},
  {Remazeilles}, {Renault}, {Ricciardi}, {Riller}, {Ristorcelli}, {Rocha},
  {Rosset}, {Roudier}, {Rowan- Robinson}, {Rubi{\~n}o-Mart{\'\i}n}, {Rusholme},
  {Sandri}, {Santos}, {Savini}, {Scott}, {Seiffert}, {Shellard}, {Spencer},
  {Starck}, {Stolyarov}, {Stompor}, {Sudiwala}, {Sunyaev}, {Sureau}, {Sutton},
  {Suur-Uski}, {Sygnet}, {Tauber}, {Tavagnacco}, {Terenzi}, {Toffolatti},
  {Tomasi}, {Tristram}, {Tucci}, {Tuovinen}, {T{\"u}rler}, {Umana},
  {Valenziano}, {Valiviita}, {Van Tent}, {Verstraete}, {Vielva}, {Villa},
  {Vittorio}, {Wade}, {Wandelt}, {Welikala}, {Ysard}, {Yvon}, {Zacchei}, \&
  {Zonca}}]{Planck_2014}
{Planck Collaboration}, {Abergel}, A., {Ade}, P.~A.~R., {et~al.} 2014, \aap,
  571, A11

\bibitem[{{Reid} {et~al.}(2016){Reid}, {Dame}, {Menten}, \&
  {Brunthaler}}]{Reid_2016}
{Reid}, M.~J., {Dame}, T.~M., {Menten}, K.~M., \& {Brunthaler}, A. 2016, \apj,
  823, 77

\bibitem[{{Reid} {et~al.}(2014){Reid}, {Menten}, {Brunthaler}, {Zheng}, {Dame},
  {Xu}, {Wu}, {Zhang}, {Sanna}, {Sato}, {Hachisuka}, {Choi}, {Immer},
  {Moscadelli}, {Rygl}, \& {Bartkiewicz}}]{Reid_2014}
{Reid}, M.~J., {Menten}, K.~M., {Brunthaler}, A., {et~al.} 2014, \apj, 783, 130

\bibitem[{{Reipurth}(2008{\natexlab{a}})}]{Northern_Handbook}
{Reipurth}, B. 2008{\natexlab{a}}, {Handbook of Star Forming Regions, Volume I:
  The Northern Sky}, Vol.~4 (San Francisco: Astronomical Society of the Pacific
  Monographs)

\bibitem[{{Reipurth}(2008{\natexlab{b}})}]{Southern_Handbook}
{Reipurth}, B. 2008{\natexlab{b}}, {Handbook of Star Forming Regions, Volume
  II: The Southern Sky}, Vol.~5 (San Francisco: Astronomical Society of the
  Pacific Monographs)

\bibitem[{{Rezaei Kh.} {et~al.}(2018){Rezaei Kh.}, {Bailer-Jones}, {Hogg}, \&
  {Schultheis}}]{Rezaei_2018}
{Rezaei Kh.}, S., {Bailer-Jones}, C. A.~L., {Hogg}, D.~W., \& {Schultheis}, M.
  2018, \aap, 618, A168

\bibitem[{Robitaille {et~al.}(2017)Robitaille, Beaumont, Qian, Borkin, \&
  Goodman}]{glueviz_2017}
Robitaille, T., Beaumont, C., Qian, P., Borkin, M., \& Goodman, A. 2017,
  Glueviz V0.13.1: Multidimensional Data Exploration, Zenodo

\bibitem[{{Sakai} {et~al.}(2019){Sakai}, {Reid}, {Menten}, {Brunthaler}, \&
  {Dame}}]{Sakai_2019}
{Sakai}, N., {Reid}, M.~J., {Menten}, K.~M., {Brunthaler}, A., \& {Dame}, T.~M.
  2019, \apj, 876, 30

\bibitem[{{Sale} \& {Magorrian}(2018)}]{Sale_2018}
{Sale}, S.~E. \& {Magorrian}, J. 2018, \mnras, 481, 494

\bibitem[{Schlafly {et~al.}(2014)Schlafly, Green, Finkbeiner, Rix, Bell,
  Burgett, Chambers, Draper, Hodapp, Kaiser, Magnier, Martin, Metcalfe, Price,
  \& Tonry}]{Schlafly_2014}
Schlafly, E.~F., Green, G., Finkbeiner, D.~P., {et~al.} 2014, \apj, 786, 29

\bibitem[{{Schlafly} {et~al.}(2018){Schlafly}, {Green}, {Lang}, {Daylan},
  {Finkbeiner}, {Lee}, {Meisner}, {Schlegel}, \& {Valdes}}]{Schlafly_2018}
{Schlafly}, E.~F., {Green}, G.~M., {Lang}, D., {et~al.} 2018, \apjs, 234, 39

\bibitem[{{Schlafly} {et~al.}(2016){Schlafly}, {Meisner}, {Stutz},
  {Kainulainen}, {Peek}, {Tchernyshyov}, {Rix}, {Finkbeiner}, {Covey}, {Green},
  {Bell}, {Burgett}, {Chambers}, {Draper}, {Flewelling}, {Hodapp}, {Kaiser},
  {Magnier}, {Martin}, {Metcalfe}, {Wainscoat}, \& {Waters}}]{Schlafly_2016}
{Schlafly}, E.~F., {Meisner}, A.~M., {Stutz}, A.~M., {et~al.} 2016, \apj, 821,
  78

\bibitem[{Shore(2002)}]{Shore_2002}
Shore, S.~N. 2002, The Tapestry of Modern Astrophysics (Wiley)

\bibitem[{{Skrutskie} {et~al.}(2006){Skrutskie}, {Cutri}, {Stiening},
  {Weinberg}, {Schneider}, {Carpenter}, {Beichman}, {Capps}, {Chester},
  {Elias}, {Huchra}, {Liebert}, {Lonsdale}, {Monet}, {Price}, {Seitzer},
  {Jarrett}, {Kirkpatrick}, {Gizis}, {Howard}, {Evans}, {Fowler}, {Fullmer},
  {Hurt}, {Light}, {Kopan}, {Marsh}, {McCallon}, {Tam}, {Van Dyk}, \&
  {Wheelock}}]{Skrutskie_2006}
{Skrutskie}, M.~F., {Cutri}, R.~M., {Stiening}, R., {et~al.} 2006, \aj, 131,
  1163

\bibitem[{{Speagle}(2019)}]{Speagle_2019}
{Speagle}, J.~S. 2019, arXiv e-prints, arXiv:1904.02180, \mnras, submitted

\bibitem[{{Wolf}(1923)}]{Wolf_1923}
{Wolf}, M. 1923, Astronomische Nachrichten, 219, 109

\bibitem[{{Xu} {et~al.}(2013){Xu}, {Li}, {Reid}, {Menten}, {Zheng},
  {Brunthaler}, {Moscadelli}, {Dame}, \& {Zhang}}]{Xu_2013}
{Xu}, Y., {Li}, J.~J., {Reid}, M.~J., {et~al.} 2013, \apj, 769, 15

\bibitem[{{Yan} {et~al.}(2019){Yan}, {Zhang}, {Xu}, {Guo}, {Macquart}, {Tang},
  \& {Walsh}}]{Yan_2019}
{Yan}, Q.-Z., {Zhang}, B., {Xu}, Y., {et~al.} 2019, \aap, 624, A6

\bibitem[{{Zhang} {et~al.}(2012){Zhang}, {Reid}, {Menten}, \&
  {Zheng}}]{Zhang_2012}
{Zhang}, B., {Reid}, M.~J., {Menten}, K.~M., \& {Zheng}, X.~W. 2012, \apj, 744,
  23

\bibitem[{{Zucker} {et~al.}(2019){Zucker}, {Speagle}, {Schlafly}, {Green},
  {Finkbeiner}, {Goodman}, \& {Alves}}]{Zucker_2019}
{Zucker}, C., {Speagle}, J.~S., {Schlafly}, E.~F., {et~al.} 2019, \apj, 879,
  125

\end{thebibliography}
\clearpage
\onecolumn

\begin{TableNotes}
\item[1] Name of the cloud associated with each sightline. We generally favor the lower density envelopes (where we can see more stars through the clouds) over the most extinguished regions
\item[2] Longitude of the sightline 
\item[3] Latitude of the sightline 
\item[4] Distance to the cloud. The first error term is the statistical uncertainty, while the second is the systematic uncertainty (estimated to be $\approx$ 5\% in distance for nearby clouds, $\approx$ 10\% in distance for faraway clouds $\gtrsim$ 1.5 kpc, and $\approx 7\%$ in distance for the NSC clouds Chamaeleon, Lupus, and Corona Australis; see Sect. \ref{details} in this work and Sect. 3.6 in \citet{Zucker_2019} for more discussion on the systematic uncertainties). We recommend the uncertainties be added in quadrature.
\item[5] Whether the sightline was taken from \citet{Zucker_2019} (Y) or is novel to this work (N)
\item[6] Whether the cloud appeared in the Northern or Southern volume of the Star Formation Handbook
\item[7] The page in the Northern or Southern volume of the Star Formation Handbook in which the cloud appeared
\end{TableNotes}

\longtab{
\setlength{\tabcolsep}{5pt}
\renewcommand{\arraystretch}{1.35}
\begin{longtable}{ccccccc}
\caption{\label{tab:distance} Molecular cloud distances}\\
\hline\hline
Cloud & l & b& Distance & Targeted in Zucker+2019? & Volume & Page \\
 & $^\circ$ & $^\circ$  & pc & pc & & \\
\hline
\endfirsthead
\caption{continued.}\\
\hline\hline
Cloud & l & b&  Distance & Targeted in Zucker+2019? & Volume & Page \\
 & $^\circ$ & $^\circ$ & pc & pc & &   \\

\hline
\endhead
\hline

\endfoot
\hline
\insertTableNotes \\  
\endlastfoot
     Aquila Rift & 21.8 & 9.2 & $278^{+11}_{-12} \pm 13$ &    N &     Northern &           18\\
     Aquila Rift & 16.7 & 11.1 & $208^{+7}_{-4} \pm 10$ &    N &     Northern &           18\\
     Aquila Rift & 18.3 & 12.7 & $254^{+4}_{-5} \pm 12$ &    N &     Northern &           18\\
     Aquila Rift & 21.5 & 11.9 & $280^{+3}_{-3} \pm 14$ &    N &     Northern &           18\\
     Aquila Rift & 16.0 & 16.6 & $163^{+3}_{-5} \pm 8$ &    N &     Northern &           18\\
        Aquila S & 38.9 & -19.1 & $123^{+3}_{-3} \pm 6$ &    Y &           -- &           --\\
        Aquila S & 39.3 & -16.8 & $128^{+3}_{-3} \pm 6$ &    Y &           -- &           --\\
        Aquila S & 37.8 & -17.5 & $135^{+2}_{-1} \pm 6$ &    Y &           -- &           --\\
        Aquila S & 36.8 & -15.1 & $143^{+3}_{-3} \pm 7$ &    Y &           -- &           --\\
             Ara & 336.7 & -2.0 & $1064^{+12}_{-15} \pm 53$ &    N &     Southern &          388\\
             Ara & 336.4 & -1.7 & $1046^{+15}_{-14} \pm 52$ &    N &     Southern &          388\\
            CB28 & 204.0 & -25.3 & $398^{+32}_{-29} \pm 19$ &    N &     Northern &          801\\
            CB29 & 205.8 & -21.6 & $374^{+10}_{-10} \pm 18$ &    N &     Northern &          801\\
            CB34 & 187.0 & -3.9 & $1322^{+13}_{-19} \pm 66$ &    N &     Northern &          869\\
         CMa OB1 & 225.4 & 0.3 & $1268^{+1}_{-4} \pm 63$ &    Y &     Southern &            1\\
         CMa OB1 & 225.0 & -0.2 & $1266^{+3}_{-2} \pm 63$ &    Y &     Southern &            1\\
         CMa OB1 & 222.9 & -1.9 & $1169^{+22}_{-6} \pm 58$ &    Y &     Southern &            1\\
         CMa OB1 & 224.5 & -0.2 & $1262^{+7}_{-13} \pm 63$ &    Y &     Southern &            1\\
      California & 162.5 & -9.5 & $436^{+11}_{-6} \pm 21$ &    Y &     Northern &          390\\
      California & 161.2 & -9.0 & $454^{+17}_{-18} \pm 22$ &    Y &     Northern &          390\\
      California & 163.8 & -7.9 & $466^{+18}_{-9} \pm 23$ &    Y &     Northern &          390\\
             Cam & 148.8 & 17.8 & $368^{+10}_{-13} \pm 18$ &    Y &     Northern &          294\\
             Cam & 144.8 & 17.8 & $220^{+12}_{-5} \pm 11$ &    Y &     Northern &          294\\
             Cam & 146.1 & 17.7 & $235^{+9}_{-8} \pm 11$ &    Y &     Northern &          294\\
             Cam & 148.4 & 17.7 & $365^{+18}_{-10} \pm 18$ &    Y &     Northern &          294\\
             Cam & 146.6 & 17.2 & $215^{+11}_{-10} \pm 10$ &    Y &     Northern &          294\\
          Carina & 286.2 & -0.2 & $2558^{+45}_{-47} \pm 255$ &    N &     Southern &          138\\
          Carina & 288.1 & -1.1 & $2439^{+95}_{-35} \pm 243$ &    N &     Southern &          138\\
          Carina & 286.3 & 0.2 & $2501^{+50}_{-41} \pm 250$ &    N &     Southern &          138\\
          Carina & 287.4 & -0.6 & $2492^{+40}_{-69} \pm 249$ &    N &     Southern &          138\\
         Cepheus & 110.1 & 17.4 & $337^{+9}_{-9} \pm 16$ &    Y &     Northern &          136\\
         Cepheus & 104.0 & 9.4 & $1045^{+24}_{-9} \pm 52$ &    Y &     Northern &          136\\
         Cepheus & 111.5 & 12.2 & $958^{+11}_{-17} \pm 47$ &    Y &     Northern &          136\\
         Cepheus & 108.3 & 17.6 & $346^{+11}_{-7} \pm 17$ &    Y &     Northern &          136\\
         Cepheus & 107.0 & 6.0 & $901^{+7}_{-6} \pm 45$ &    Y &     Northern &          136\\
         Cepheus & 103.7 & 11.4 & $867^{+4}_{-9} \pm 43$ &    Y &     Northern &          136\\
         Cepheus & 104.0 & 14.5 & $341^{+18}_{-14} \pm 17$ &    Y &     Northern &          136\\
         Cepheus & 107.7 & 5.9 & $850^{+16}_{-26} \pm 42$ &    Y &     Northern &          136\\
         Cepheus & 111.5 & 20.8 & $331^{+11}_{-20} \pm 16$ &    Y &     Northern &          136\\
         Cepheus & 114.6 & 16.5 & $346^{+4}_{-5} \pm 17$ &    Y &     Northern &          136\\
         Cepheus & 109.6 & 16.9 & $344^{+6}_{-6} \pm 17$ &    Y &     Northern &          136\\
         Cepheus & 113.5 & 15.9 & $336^{+3}_{-4} \pm 16$ &    Y &     Northern &          136\\
         Cepheus & 108.2 & 5.5 & $891^{+10}_{-8} \pm 44$ &    Y &     Northern &          136\\
         Cepheus & 107.7 & 12.4 & $961^{+7}_{-4} \pm 48$ &    Y &     Northern &          136\\
         Cepheus & 115.3 & 17.6 & $358^{+6}_{-6} \pm 17$ &    Y &     Northern &          136\\
         Cepheus & 112.8 & 20.8 & $375^{+11}_{-13} \pm 18$ &    Y &     Northern &          136\\
         Cepheus & 111.8 & 20.3 & $364^{+6}_{-5} \pm 18$ &    Y &     Northern &          136\\
         Cepheus & 106.4 & 17.7 & $377^{+5}_{-5} \pm 18$ &    Y &     Northern &          136\\
         Cepheus & 109.0 & 7.7 & $816^{+24}_{-15} \pm 40$ &    Y &     Northern &          136\\
         Cepheus & 112.8 & 16.5 & $344^{+7}_{-9} \pm 17$ &    Y &     Northern &          136\\
         Cepheus & 103.5 & 13.5 & $359^{+5}_{-4} \pm 17$ &    Y &     Northern &          136\\
         Cepheus & 107.0 & 9.4 & $986^{+9}_{-9} \pm 49$ &    Y &     Northern &          136\\
         Cepheus & 110.7 & 12.6 & $989^{+4}_{-23} \pm 49$ &    Y &     Northern &          136\\
         Cepheus & 108.4 & 18.6 & $332^{+31}_{-17} \pm 16$ &    Y &     Northern &          136\\
         Cepheus & 109.6 & 6.8 & $881^{+14}_{-16} \pm 44$ &    Y &     Northern &          136\\
         Cepheus & 108.3 & 12.4 & $915^{+3}_{-3} \pm 45$ &    Y &     Northern &          136\\
         Cepheus & 105.9 & 13.8 & $951^{+6}_{-12} \pm 47$ &    Y &     Northern &          136\\
         Cepheus & 116.1 & 20.2 & $349^{+8}_{-8} \pm 17$ &    Y &     Northern &          136\\
      Chamaeleon & 303.3 & -14.2 & $190^{+4}_{-4} \pm 13$ &    N &     Southern &          169\\
      Chamaeleon & 297.5 & -15.3 & $210^{+15}_{-11} \pm 14$ &    N &     Southern &          169\\
      Chamaeleon & 303.0 & -16.7 & $161^{+6}_{-8} \pm 11$ &    N &     Southern &          169\\
        Circinus & 318.4 & -3.2 & $683^{+1}_{-3} \pm 34$ &    N &     Southern &          285\\
        Circinus & 316.9 & -3.9 & $675^{+1}_{-3} \pm 33$ &    N &     Southern &          285\\
        Coalsack & 302.9 & -2.6 & $188^{+8}_{-6} \pm 9$ &    N &     Southern &          222\\
        Coalsack & 301.4 & -2.6 & $182^{+5}_{-6} \pm 9$ &    N &     Southern &          222\\
        Coalsack & 301.4 & 3.1 & $192^{+2}_{-2} \pm 9$ &    N &     Southern &          222\\
        Coalsack & 302.8 & 1.8 & $187^{+12}_{-9} \pm 9$ &    N &     Southern &          222\\
        Coalsack & 302.9 & 3.1 & $191^{+4}_{-5} \pm 9$ &    N &     Southern &          222\\
        Coalsack & 300.0 & 3.1 & $193^{+2}_{-8} \pm 9$ &    N &     Southern &          222\\
Corona Australis & 0.8 & -20.1 & $155^{+5}_{-6} \pm 10$ &    N &     Southern &          735\\
Corona Australis & 359.5 & -17.8 & $147^{+5}_{-5} \pm 10$ &    N &     Southern &          735\\
Corona Australis & 359.5 & -21.0 & $165^{+3}_{-4} \pm 11$ &    N &     Southern &          735\\
         CygnusX & 82.9 & 0.7 & $1272^{+13}_{-12} \pm 63$ &    N &     Northern &           36\\
         CygnusX & 76.8 & 2.2 & $1622^{+15}_{-25} \pm 162$ &    N &     Northern &           36\\
         CygnusX & 77.2 & 2.1 & $1309^{+24}_{-10} \pm 65$ &    N &     Northern &           36\\
         CygnusX & 80.3 & -2.4 & $1441^{+7}_{-9} \pm 72$ &    N &     Northern &           36\\
         CygnusX & 80.5 & 1.1 & $1214^{+21}_{-9} \pm 60$ &    N &     Northern &           36\\
         CygnusX & 79.0 & 3.7 & $1507^{+23}_{-22} \pm 150$ &    N &     Northern &           36\\
         CygnusX & 77.7 & 1.3 & $898^{+12}_{-6} \pm 44$ &    N &     Northern &           36\\
         CygnusX & 80.0 & -0.7 & $991^{+17}_{-16} \pm 49$ &    N &     Northern &           36\\
         CygnusX & 79.1 & 3.0 & $1003^{+8}_{-12} \pm 50$ &    N &     Northern &           36\\
         CygnusX & 80.3 & 2.9 & $973^{+4}_{-8} \pm 48$ &    N &     Northern &           36\\
         CygnusX & 80.2 & 0.1 & $1226^{+20}_{-10} \pm 61$ &    N &     Northern &           36\\
         CygnusX & 82.3 & 1.0 & $761^{+14}_{-10} \pm 38$ &    N &     Northern &           36\\
         CygnusX & 78.7 & 0.6 & $919^{+14}_{-36} \pm 45$ &    N &     Northern &           36\\
           Draco & 89.5 & 38.4 & $481^{+51}_{-45} \pm 24$ &    Y &     Southern &          813\\
            GGD4 & 184.3 & -4.2 & $1349^{+34}_{-22} \pm 67$ &    N &     Northern &          869\\
            GGD4 & 185.1 & -4.3 & $1396^{+23}_{-22} \pm 69$ &    N &     Northern &          869\\
         Gem OB1 & 189.9 & -0.3 & $1815^{+11}_{-32} \pm 181$ &    N &     Northern &          869\\
         Gem OB1 & 189.5 & 0.7 & $1864^{+12}_{-16} \pm 186$ &    N &     Northern &          869\\
         Gem OB1 & 190.9 & 0.0 & $1865^{+12}_{-8} \pm 186$ &    N &     Northern &          869\\
         Gem OB1 & 188.7 & 1.0 & $1971^{+7}_{-10} \pm 197$ &    N &     Northern &          869\\
         Gem OB1 & 193.0 & 0.6 & $1726^{+25}_{-9} \pm 172$ &    N &     Northern &          869\\
        Hercules & 45.1 & 8.9 & $223^{+3}_{-2} \pm 11$ &    Y &           -- &           --\\
        Hercules & 44.1 & 8.6 & $223^{+3}_{-2} \pm 11$ &    Y &           -- &           --\\
        Hercules & 42.8 & 7.9 & $230^{+4}_{-2} \pm 11$ &    Y &           -- &           --\\
          IC1396 & 98.9 & 4.0 & $916^{+21}_{-13} \pm 45$ &    N &     Northern &          136\\
          IC1396 & 100.1 & 4.2 & $905^{+21}_{-17} \pm 45$ &    N &     Northern &          136\\
          IC1396 & 100.4 & 3.4 & $941^{+29}_{-29} \pm 47$ &    N &     Northern &          136\\
          IC1396 & 99.1 & 4.7 & $909^{+18}_{-25} \pm 45$ &    N &     Northern &          136\\
          IC2118 & 206.4 & -26.0 & $328^{+15}_{-20} \pm 16$ &    N &     Northern &          459\\
          IC2118 & 207.3 & -27.2 & $273^{+8}_{-11} \pm 13$ &    N &     Northern &          459\\
          IC2118 & 206.9 & -26.6 & $283^{+16}_{-30} \pm 14$ &    N &     Northern &          459\\
          IC2944 & 294.1 & -1.6 & $2363^{+39}_{-17} \pm 236$ &    N &     Southern &          213\\
          IC2944 & 294.9 & -1.5 & $2452^{+14}_{-22} \pm 245$ &    N &     Southern &          213\\
          IC2944 & 294.6 & -2.0 & $2342^{+20}_{-41} \pm 234$ &    N &     Southern &          213\\
           IC443 & 189.2 & 3.2 & $1629^{+26}_{-19} \pm 162$ &    N &     Northern &          869\\
           IC443 & 189.2 & 4.7 & $1593^{+26}_{-15} \pm 159$ &    N &     Northern &          869\\
           IC443 & 189.6 & 4.0 & $1558^{+6}_{-8} \pm 155$ &    N &     Northern &          869\\
           IC443 & 189.1 & 4.1 & $1588^{+50}_{-30} \pm 158$ &    N &     Northern &          869\\
          IC5146 & 93.7 & -4.6 & $774^{+13}_{-16} \pm 38$ &    N &     Northern &          108\\
          IC5146 & 93.4 & -4.2 & $792^{+13}_{-15} \pm 39$ &    N &     Northern &          108\\
          IC5146 & 94.0 & -4.9 & $730^{+19}_{-25} \pm 36$ &    N &     Northern &          108\\
          IC5146 & 94.4 & -5.5 & $751^{+10}_{-8} \pm 37$ &    N &     Northern &          108\\
           L1228 & 111.8 & 20.2 & $366^{+6}_{-5} \pm 18$ &    N &     Northern &          136\\
          L1228D & 112.3 & 13.8 & $491^{+20}_{-160} \pm 24$ &    N &     Northern &          136\\
           L1251 & 114.6 & 14.5 & $351^{+10}_{-6} \pm 17$ &    N &     Northern &          136\\
           L1265 & 115.9 & -2.0 & $344^{+34}_{-31} \pm 17$ &    N &     Northern &          240\\
           L1293 & 121.7 & 0.1 & $1083^{+12}_{-12} \pm 54$ &    N &     Northern &          240\\
           L1302 & 122.0 & -1.4 & $906^{+13}_{-7} \pm 45$ &    N &     Northern &          240\\
           L1306 & 125.6 & -0.6 & $903^{+15}_{-13} \pm 45$ &    N &     Northern &          240\\
           L1306 & 126.8 & -0.8 & $941^{+17}_{-20} \pm 47$ &    N &     Northern &          240\\
           L1307 & 124.3 & 3.3 & $834^{+6}_{-5} \pm 41$ &    N &     Northern &          240\\
           L1307 & 124.6 & 2.6 & $902^{+20}_{-10} \pm 45$ &    N &     Northern &          240\\
           L1333 & 128.9 & 13.7 & $283^{+3}_{-3} \pm 14$ &    N &     Northern &          240\\
           L1335 & 128.9 & 4.3 & $647^{+26}_{-18} \pm 32$ &    N &     Northern &          240\\
           L1340 & 130.0 & 11.5 & $858^{+10}_{-12} \pm 42$ &    N &     Northern &          240\\
           L1355 & 132.8 & 8.9 & $948^{+34}_{-16} \pm 47$ &    N &     Northern &          240\\
           L1355 & 133.6 & 9.3 & $924^{+13}_{-14} \pm 46$ &    N &     Northern &          240\\
           L1616 & 203.5 & -24.8 & $392^{+8}_{-7} \pm 19$ &    N &     Northern &          801\\
           L1617 & 203.5 & -12.0 & $414^{+5}_{-9} \pm 20$ &    N &     Northern &          782\\
           L1622 & 204.7 & -11.8 & $418^{+17}_{-17} \pm 20$ &    N &     Northern &          782\\
           L1634 & 207.6 & -23.2 & $389^{+24}_{-12} \pm 19$ &    N &     Northern &          801\\
           L1634 & 207.6 & -22.8 & $364^{+19}_{-13} \pm 18$ &    N &     Northern &          801\\
            L291 & 10.7 & -2.8 & $1439^{+17}_{-19} \pm 71$ &    N &     Southern &          578\\
            L291 & 11.3 & -2.1 & $1336^{+12}_{-22} \pm 66$ &    N &     Southern &          578\\
            L291 & 11.4 & -2.7 & $1348^{+22}_{-24} \pm 67$ &    N &     Southern &          578\\
            L379 & 16.9 & -2.2 & $2406^{+35}_{-52} \pm 240$ &    N &           -- &           --\\
            L379 & 17.1 & -2.9 & $1890^{+99}_{-29} \pm 189$ &    N &           -- &           --\\
            L379 & 16.8 & -2.7 & $2061^{+13}_{-20} \pm 206$ &    N &           -- &           --\\
            L977 & 89.5 & 2.0 & $660^{+15}_{-12} \pm 33$ &    N &     Northern &           36\\
            L977 & 89.8 & 2.2 & $642^{+16}_{-16} \pm 32$ &    N &     Northern &           36\\
            L988 & 90.7 & 2.4 & $627^{+6}_{-6} \pm 31$ &    N &     Northern &           36\\
            L988 & 90.2 & 2.3 & $612^{+7}_{-6} \pm 30$ &    N &     Northern &           36\\
          LBN906 & 202.2 & -31.4 & $287^{+5}_{-4} \pm 14$ &    N &     Northern &          801\\
          LBN917 & 203.7 & -30.3 & $232^{+7}_{-6} \pm 11$ &    N &     Northern &          801\\
          LBN942 & 205.6 & -22.0 & $257^{+6}_{-6} \pm 12$ &    N &     Northern &          801\\
          LBN968 & 208.4 & -28.4 & $319^{+14}_{-10} \pm 15$ &    N &     Northern &          801\\
          LBN969 & 208.4 & -22.2 & $274^{+7}_{-5} \pm 13$ &    N &     Northern &          801\\
          LBN991 & 213.6 & -28.9 & $408^{+20}_{-16} \pm 20$ &    N &     Northern &          801\\
         Lacerta & 96.1 & -10.2 & $504^{+7}_{-5} \pm 25$ &    Y &     Northern &          124\\
         Lacerta & 95.8 & -11.5 & $473^{+5}_{-4} \pm 23$ &    Y &     Northern &          124\\
          Lagoon & 6.9 & -2.2 & $1325^{+8}_{-7} \pm 66$ &    N &     Southern &          533\\
          Lagoon & 7.3 & -2.4 & $1220^{+11}_{-9} \pm 61$ &    N &     Southern &          533\\
           Lupus & 347.3 & 6.6 & $239^{+48}_{-51} \pm 16$ &    N &     Southern &          295\\
           Lupus & 338.7 & 17.4 & $160^{+5}_{-6} \pm 11$ &    N &     Southern &          295\\
           Lupus & 339.0 & 16.3 & $151^{+13}_{-11} \pm 10$ &    N &     Southern &          295\\
           Lupus & 341.1 & 9.8 & $197^{+5}_{-5} \pm 13$ &    N &     Southern &          295\\
           Lupus & 341.1 & 6.4 & $108^{+52}_{-33} \pm 7$ &    N &     Southern &          295\\
           Lupus & 339.0 & 14.9 & $156^{+4}_{-6} \pm 10$ &    N &     Southern &          295\\
             M16 & 17.9 & 1.0 & $1640^{+22}_{-29} \pm 164$ &    N &     Southern &          599\\
             M16 & 17.5 & 1.2 & $1739^{+23}_{-19} \pm 173$ &    N &     Southern &          599\\
             M17 & 15.5 & -0.8 & $1488^{+24}_{-14} \pm 74$ &    N &     Southern &          624\\
             M17 & 15.2 & -0.3 & $1509^{+22}_{-24} \pm 150$ &    N &     Southern &          624\\
             M17 & 15.3 & -1.1 & $1574^{+17}_{-34} \pm 157$ &    N &     Southern &          624\\
             M20 & 7.4 & -0.5 & $1253^{+10}_{-11} \pm 62$ &    N &     Southern &          509\\
             M20 & 6.7 & -0.5 & $1234^{+19}_{-23} \pm 61$ &    N &     Southern &          509\\
             M20 & 6.6 & -0.1 & $1184^{+12}_{-8} \pm 59$ &    N &     Southern &          509\\
             M20 & 7.0 & 0.1 & $1186^{+10}_{-6} \pm 59$ &    N &     Southern &          509\\
       Maddalena & 216.5 & -2.5 & $2110^{+10}_{-5} \pm 211$ &    Y &           -- &           --\\
       Maddalena & 217.1 & 0.4 & $1888^{+22}_{-13} \pm 188$ &    Y &           -- &           --\\
       Maddalena & 216.4 & 0.1 & $2099^{+16}_{-10} \pm 209$ &    Y &           -- &           --\\
       Maddalena & 216.8 & -2.2 & $2113^{+17}_{-9} \pm 211$ &    Y &           -- &           --\\
Mon OB1 (NGC2264) & 202.1 & 2.7 & $771^{+14}_{-6} \pm 38$ &    N &     Northern &          966\\
Mon OB1 (NGC2264) & 202.8 & 2.3 & $759^{+10}_{-26} \pm 37$ &    N &     Northern &          966\\
Mon OB1 (NGC2264) & 203.1 & 1.8 & $780^{+16}_{-12} \pm 39$ &    N &     Northern &          966\\
         Mon OB1 (NGC2264)& 201.2 & 1.0 & $715^{+46}_{-7} \pm 35$ &    Y &     Northern &          966\\
         Mon OB1 (NGC2264) & 200.4 & 0.8 & $719^{+11}_{-5} \pm 35$ &    Y &     Northern &          966\\
         Mon OB1 (NGC2264) & 201.4 & 1.1 & $748^{+10}_{-11} \pm 37$ &    Y &     Northern &          966\\
        Mon R2 & 213.0 & -12.5 & $799^{+5}_{-3} \pm 39$ &    N &     Northern &          899\\
          Mon R2 & 219.3 & -9.5 & $923^{+10}_{-14} \pm 46$ &    Y &     Northern &          899\\
          Mon R2 & 215.3 & -12.9 & $767^{+13}_{-17} \pm 38$ &    Y &     Northern &          899\\
          Mon R2 & 219.2 & -7.7 & $943^{+36}_{-5} \pm 47$ &    Y &     Northern &          899\\
          Mon R2 & 220.9 & -8.3 & $915^{+5}_{-4} \pm 45$ &    Y &     Northern &          899\\
          Mon R2 & 213.9 & -11.9 & $788^{+12}_{-15} \pm 39$ &    Y &     Northern &          899\\
         NGC2362 & 239.5 & -4.9 & $1317^{+7}_{-6} \pm 65$ &    N &     Southern &           26\\
         NGC2362 & 237.2 & -4.9 & $1173^{+17}_{-19} \pm 58$ &    N &     Southern &           26\\
         NGC2362 & 238.7 & -4.2 & $1358^{+10}_{-8} \pm 67$ &    N &     Southern &           26\\
         NGC6604 & 17.8 & 2.2 & $1352^{+7}_{-21} \pm 67$ &    N &     Southern &          590\\
         NGC6604 & 18.1 & 1.9 & $1524^{+19}_{-35} \pm 152$ &    N &     Southern &          590\\
         NGC6604 & 18.2 & 2.5 & $1334^{+8}_{-7} \pm 66$ &    N &     Southern &          590\\
           Norma & 338.8 & 1.8 & $721^{+56}_{-46} \pm 36$ &    N &     Southern &          381\\
   North America & 85.8 & -2.2 & $834^{+18}_{-25} \pm 41$ &    N &     Northern &           36\\
   North America & 84.8 & -1.2 & $878^{+29}_{-23} \pm 43$ &    N &     Northern &           36\\
   North America & 85.3 & -0.8 & $784^{+118}_{-31} \pm 39$ &    N &     Northern &           36\\
   North America & 86.6 & -2.1 & $731^{+40}_{-21} \pm 36$ &    N &     Northern &           36\\
   North America & 84.6 & 0.1 & $792^{+8}_{-8} \pm 39$ &    N &     Northern &           36\\
   North America & 82.8 & -2.1 & $809^{+12}_{-13} \pm 40$ &    N &     Northern &           36\\
   North America & 83.7 & -2.1 & $781^{+11}_{-13} \pm 39$ &    N &     Northern &           36\\
   North America & 81.7 & -1.7 & $852^{+17}_{-18} \pm 42$ &    N &     Northern &           36\\
   North America & 84.2 & -1.3 & $811^{+7}_{-14} \pm 40$ &    N &     Northern &           36\\
   North America & 84.8 & -1.7 & $821^{+13}_{-11} \pm 41$ &    N &     Northern &           36\\
Northern Coalsack & 91.3 & 4.2 & $577^{+8}_{-7} \pm 28$ &    N &     Northern &           36\\
Northern Coalsack & 92.6 & 3.5 & $561^{+13}_{-13} \pm 28$ &    N &     Northern &           36\\
Northern Coalsack & 92.2 & 4.3 & $547^{+6}_{-8} \pm 27$ &    N &     Northern &           36\\
       Ophiuchus & 352.7 & 15.4 & $139^{+3}_{-2} \pm 6$ &    Y &     Southern &          351\\
       Ophiuchus & 355.2 & 16.0 & $128^{+3}_{-2} \pm 6$ &    Y &     Southern &          351\\
       Ophiuchus & 357.1 & 15.7 & $118^{+5}_{-4} \pm 5$ &    Y &     Southern &          351\\
 Ophiuchus (Arc) & 349.9 & 16.6 & $167^{+5}_{-5} \pm 8$ &    N &     Southern &          351\\
 Ophiuchus (Arc) & 349.3 & 14.9 & $155^{+2}_{-3} \pm 7$ &    N &     Southern &          351\\
 Ophiuchus (Arc) & 352.4 & 18.3 & $130^{+2}_{-3} \pm 6$ &    N &     Southern &          351\\
 Ophiuchus (B44) & 359.2 & 12.0 & $149^{+5}_{-4} \pm 7$ &    N &     Southern &          351\\
 Ophiuchus (B44) & 357.1 & 13.1 & $145^{+5}_{-5} \pm 7$ &    N &     Southern &          351\\
 Ophiuchus (B44) & 354.5 & 15.0 & $154^{+4}_{-4} \pm 7$ &    N &     Southern &          351\\
 Ophiuchus (B45) & 358.6 & 15.3 & $139^{+4}_{-9} \pm 6$ &    N &     Southern &          351\\
 Ophiuchus (B45) & 357.1 & 15.6 & $142^{+6}_{-5} \pm 7$ &    N &     Southern &          351\\
 Ophiuchus (B45) & 355.6 & 16.1 & $132^{+3}_{-2} \pm 6$ &    N &     Southern &          351\\
Ophiuchus (L1688) & 353.2 & 16.6 & $139^{+3}_{-4} \pm 6$ &    N &     Southern &          351\\
Ophiuchus (North) & 8.4 & 22.0 & $109^{+8}_{-5} \pm 5$ &    N &     Southern &          351\\
Ophiuchus (North) & 4.2 & 18.2 & $151^{+3}_{-5} \pm 7$ &    N &     Southern &          351\\
Ophiuchus (North) & 6.5 & 20.4 & $134^{+10}_{-11} \pm 6$ &    N &     Southern &          351\\
           Orion & 206.4 & -15.4 & $433^{+27}_{-22} \pm 21$ &    N &     Northern & 459, 483, 544, 662, 693\\
           Orion & 209.0 & -20.1 & $394^{+10}_{-10} \pm 19$ &    Y &     Northern & 459, 483, 544, 662, 693\\
           Orion & 209.1 & -19.9 & $445^{+25}_{-20} \pm 22$ &    Y &     Northern & 459, 483, 544, 662, 693\\
           Orion & 212.2 & -18.6 & $473^{+7}_{-6} \pm 23$ &    Y &     Northern & 459, 483, 544, 662, 693\\
           Orion & 202.0 & -13.3 & $481^{+10}_{-14} \pm 24$ &    Y &     Northern & 459, 483, 544, 662, 693\\
           Orion & 205.7 & -14.8 & $436^{+24}_{-23} \pm 21$ &    N &     Northern & 459, 483, 544, 662, 693\\
           Orion & 212.4 & -19.9 & $415^{+10}_{-16} \pm 20$ &    Y &     Northern & 459, 483, 544, 662, 693\\
           Orion & 208.4 & -19.6 & $399^{+14}_{-7} \pm 19$ &    Y &     Northern & 459, 483, 544, 662, 693\\
           Orion & 207.9 & -16.8 & $411^{+9}_{-14} \pm 20$ &    Y &     Northern & 459, 483, 544, 662, 693\\
           Orion & 214.7 & -19.0 & $416^{+4}_{-6} \pm 20$ &    Y &     Northern & 459, 483, 544, 662, 693\\
           Orion & 209.8 & -19.5 & $438^{+15}_{-27} \pm 21$ &    Y &     Northern & 459, 483, 544, 662, 693\\
           Orion & 202.0 & -14.0 & $399^{+4}_{-2} \pm 19$ &    Y &     Northern & 459, 483, 544, 662, 693\\
           Orion & 204.7 & -19.2 & $418^{+15}_{-18} \pm 20$ &    Y &     Northern & 459, 483, 544, 662, 693\\
           Orion & 207.4 & -16.0 & $451^{+8}_{-3} \pm 22$ &    N &     Northern & 459, 483, 544, 662, 693\\
           Orion & 204.8 & -13.3 & $415^{+4}_{-5} \pm 20$ &    N &     Northern & 459, 483, 544, 662, 693\\
           Orion & 212.4 & -17.3 & $522^{+20}_{-54} \pm 26$ &    Y &     Northern & 459, 483, 544, 662, 693\\
           Orion & 201.3 & -13.8 & $420^{+6}_{-10} \pm 21$ &    Y &     Northern & 459, 483, 544, 662, 693\\
       Orion Lam & 192.3 & -8.9 & $406^{+16}_{-17} \pm 20$ &    Y &     Northern &          757\\
       Orion Lam & 196.7 & -16.1 & $426^{+10}_{-7} \pm 21$ &    Y &     Northern &          757\\
       Orion Lam & 195.5 & -13.7 & $399^{+14}_{-12} \pm 19$ &    Y &     Northern &          757\\
       Orion Lam & 199.6 & -11.9 & $393^{+8}_{-4} \pm 19$ &    Y &     Northern &          757\\
       Orion Lam & 194.8 & -12.1 & $423^{+12}_{-7} \pm 21$ &    Y &     Northern &          757\\
       Orion Lam & 194.7 & -10.1 & $425^{+25}_{-5} \pm 21$ &    Y &     Northern &          757\\
       Orion Lam & 196.9 & -8.2 & $394^{+13}_{-8} \pm 19$ &    Y &     Northern &          757\\
         Pegasus & 104.2 & -31.7 & $292^{+15}_{-18} \pm 14$ &    Y &           -- &           --\\
         Pegasus & 88.8 & -41.3 & $238^{+10}_{-9} \pm 11$ &    Y &           -- &           --\\
         Pegasus & 92.2 & -34.7 & $258^{+31}_{-99} \pm 12$ &    Y &           -- &           --\\
         Pegasus & 95.3 & -35.7 & $257^{+19}_{-15} \pm 12$ &    Y &           -- &           --\\
         Pegasus & 105.6 & -30.6 & $256^{+15}_{-15} \pm 12$ &    Y &           -- &           --\\
         Perseus & 159.9 & -18.1 & $305^{+14}_{-20} \pm 15$ &    Y &     Northern & 308, 346, 372\\
         Perseus & 158.6 & -19.9 & $291^{+15}_{-6} \pm 14$ &    Y &     Northern & 308, 346, 372\\
         Perseus & 158.5 & -22.1 & $243^{+12}_{-13} \pm 12$ &    Y &     Northern & 308, 346, 372\\
         Perseus & 159.3 & -20.6 & $276^{+12}_{-8} \pm 13$ &    Y &     Northern & 308, 346, 372\\
         Perseus & 159.7 & -19.7 & $347^{+22}_{-25} \pm 17$ &    Y &     Northern & 308, 346, 372\\
         Perseus & 159.9 & -18.9 & $279^{+18}_{-13} \pm 13$ &    Y &     Northern & 308, 346, 372\\
         Perseus & 159.4 & -21.3 & $234^{+39}_{-70} \pm 11$ &    Y &     Northern & 308, 346, 372\\
         Perseus & 157.8 & -22.8 & $264^{+11}_{-7} \pm 13$ &    Y &     Northern & 308, 346, 372\\
         Perseus & 160.8 & -17.0 & $276^{+7}_{-4} \pm 13$ &    Y &     Northern & 308, 346, 372\\
         Perseus & 160.8 & -18.7 & $285^{+18}_{-15} \pm 14$ &    Y &     Northern & 308, 346, 372\\
         Perseus & 159.1 & -21.1 & $291^{+15}_{-14} \pm 14$ &    Y &     Northern & 308, 346, 372\\
         Perseus & 157.7 & -21.4 & $240^{+11}_{-12} \pm 12$ &    Y &     Northern & 308, 346, 372\\
         Perseus & 158.2 & -20.9 & $287^{+8}_{-8} \pm 14$ &    Y &     Northern & 308, 346, 372\\
         Perseus & 157.5 & -17.9 & $287^{+8}_{-8} \pm 14$ &    Y &     Northern & 308, 346, 372\\
         Perseus & 160.4 & -17.2 & $284^{+14}_{-16} \pm 14$ &    Y &     Northern & 308, 346, 372\\
         Perseus & 160.0 & -17.6 & $331^{+15}_{-10} \pm 16$ &    Y &     Northern & 308, 346, 372\\
         Perseus & 160.4 & -16.7 & $256^{+21}_{-44} \pm 12$ &    Y &     Northern & 308, 346, 372\\
         Perseus & 160.7 & -16.3 & $296^{+10}_{-7} \pm 14$ &    Y &     Northern & 308, 346, 372\\
      Pipe (B59) & 356.9 & 7.3 & $180^{+5}_{-7} \pm 9$ &    N &     Southern &          415\\
         Polaris & 123.5 & 37.9 & $472^{+35}_{-38} \pm 23$ &    Y &     Southern &          813\\
         Polaris & 129.5 & 17.3 & $341^{+20}_{-19} \pm 17$ &    Y &     Southern &          813\\
         Polaris & 126.3 & 21.2 & $343^{+6}_{-10} \pm 17$ &    Y &     Southern &          813\\
           RCW38 & 268.0 & -1.2 & $1595^{+45}_{-17} \pm 159$ &    N &     Southern &          124\\
           RCW38 & 267.7 & -1.2 & $1650^{+17}_{-25} \pm 165$ &    N &     Southern &          124\\
         Rosette & 205.2 & -2.6 & $1413^{+6}_{-6} \pm 70$ &    Y &     Northern &          928\\
         Rosette & 206.8 & -1.2 & $1356^{+7}_{-15} \pm 67$ &    Y &     Northern &          928\\
         Rosette & 207.8 & -2.1 & $1261^{+20}_{-13} \pm 63$ &    Y &     Northern &          928\\
            S106 & 76.0 & -0.7 & $1091^{+22}_{-19} \pm 54$ &    N &     Northern &           90\\
         Serpens & 29.6 & 3.9 & $501^{+11}_{-9} \pm 25$ &    N &     Southern &      683,693\\
         Serpens & 29.2 & 4.1 & $556^{+18}_{-23} \pm 27$ &    N &     Southern &      683,693\\
         Serpens & 28.5 & 3.0 & $489^{+19}_{-13} \pm 24$ &    N &     Southern &      683,693\\
         Serpens & 28.1 & 3.6 & $439^{+13}_{-23} \pm 21$ &    N &     Southern &      683,693\\
   Serpens (Low) & 31.8 & 2.6 & $494^{+17}_{-15} \pm 24$ &    N &     Southern &      683,693\\
   Serpens (Low) & 30.1 & 2.7 & $466^{+10}_{-18} \pm 23$ &    N &     Southern &      683,693\\
   Serpens (Low) & 31.9 & 3.0 & $487^{+18}_{-12} \pm 24$ &    N &     Southern &      683,693\\
   Serpens (Low) & 32.9 & 2.7 & $526^{+14}_{-14} \pm 26$ &    N &     Southern &      683,693\\
  Serpens (Main) & 30.3 & 5.2 & $490^{+14}_{-11} \pm 24$ &    N &     Southern &      683,693\\
  Serpens (Main) & 31.2 & 5.2 & $425^{+12}_{-16} \pm 21$ &    N &     Southern &      683,693\\
   Serpens (W40) & 28.8 & 3.5 & $487^{+27}_{-23} \pm 24$ &    N &     Southern &      683,693\\
     Serpens OB2 & 18.8 & 1.2 & $1569^{+88}_{-28} \pm 156$ &    N &     Southern &          590\\
     Serpens OB2 & 18.0 & 1.6 & $1577^{+18}_{-21} \pm 157$ &    N &     Southern &          590\\
     Serpens OB2 & 18.4 & 1.3 & $1611^{+23}_{-15} \pm 161$ &    N &     Southern &          590\\
         Sh2-231 & 173.0 & 2.4 & $1616^{+11}_{-28} \pm 161$ &    N &     Northern &          869\\
         Sh2-232 & 173.5 & 2.9 & $1713^{+18}_{-18} \pm 171$ &    N &     Northern &          869\\
          Spider & 134.8 & 40.5 & $369^{+19}_{-22} \pm 18$ &    Y &           -- &           --\\
          Taurus & 173.5 & -14.2 & $147^{+10}_{-15} \pm 7$ &    Y &     Northern &          405\\
          Taurus & 171.6 & -15.8 & $130^{+9}_{-8} \pm 6$ &    Y &     Northern &          405\\
          Taurus & 175.8 & -12.9 & $156^{+3}_{-2} \pm 7$ &    Y &     Northern &          405\\
          Taurus & 172.2 & -14.6 & $137^{+4}_{-2} \pm 6$ &    Y &     Northern &          405\\
          Taurus & 170.2 & -12.3 & $170^{+10}_{-5} \pm 8$ &    Y &     Northern &          405\\
          Taurus & 174.5 & -15.6 & $159^{+3}_{-3} \pm 7$ &    N &     Northern &          405\\
          Taurus & 171.7 & -17.2 & $149^{+1}_{-2} \pm 7$ &    N &     Northern &          405\\
          Taurus & 166.2 & -16.6 & $138^{+1}_{-2} \pm 6$ &    Y &     Northern &          405\\
          Taurus & 171.4 & -13.5 & $154^{+4}_{-3} \pm 7$ &    Y &     Northern &          405\\
          Taurus & 169.9 & -19.2 & $129^{+3}_{-1} \pm 6$ &    N &     Northern &          405\\
      Ursa Major & 158.5 & 35.2 & $352^{+11}_{-14} \pm 17$ &    Y &     Southern &          813\\
      Ursa Major & 143.4 & 38.5 & $408^{+8}_{-4} \pm 20$ &    Y &     Southern &          813\\
      Ursa Major & 146.9 & 40.7 & $330^{+18}_{-20} \pm 16$ &    Y &     Southern &          813\\
      Ursa Major & 153.5 & 36.7 & $352^{+19}_{-17} \pm 17$ &    Y &     Southern &          813\\
          Vela C & 266.1 & 1.4 & $866^{+21}_{-8} \pm 43$ &    N &     Southern &           43\\
          Vela C & 265.3 & 1.2 & $947^{+8}_{-7} \pm 47$ &    N &     Southern &           43\\
          Vela C & 265.3 & 1.8 & $878^{+6}_{-7} \pm 43$ &    N &     Southern &           43\\
          Vela C & 264.7 & 1.4 & $931^{+9}_{-10} \pm 46$ &    N &     Southern &           43\\
          Vela C & 264.3 & 2.5 & $965^{+15}_{-5} \pm 48$ &    N &     Southern &           43\\
              W3 & 133.3 & 1.1 & $1873^{+11}_{-20} \pm 187$ &    N &     Northern &          264\\
              W3 & 133.7 & 1.3 & $2184^{+19}_{-28} \pm 218$ &    N &     Northern &          264\\
              W3 & 133.3 & 0.5 & $1659^{+24}_{-31} \pm 165$ &    N &     Northern &          264\\
              W4 & 135.6 & 1.3 & $1825^{+31}_{-85} \pm 182$ &    N &     Northern &          264\\
              W4 & 135.3 & 1.0 & $1647^{+20}_{-9} \pm 164$ &    N &     Northern &          264\\
              W4 & 135.6 & 0.2 & $1451^{+13}_{-61} \pm 72$ &    N &     Northern &          264\\
              W4 & 135.6 & 1.2 & $1755^{+39}_{-27} \pm 175$ &    N &     Northern &          264\\
              W5 & 136.5 & 1.2 & $2026^{+24}_{-36} \pm 202$ &    N &     Northern &          264\\
              W5 & 137.0 & 1.4 & $2077^{+16}_{-18} \pm 207$ &    N &     Northern &          264\\
              W5 & 137.8 & 1.5 & $1962^{+33}_{-28} \pm 196$ &    N &     Northern &          264\\
              W5 & 138.0 & 0.9 & $1739^{+9}_{-15} \pm 173$ &    N &     Northern &          264\\
              W5 & 136.9 & 1.0 & $2103^{+13}_{-15} \pm 210$ &    N &     Northern &          264\\

\end{longtable}}

\clearpage
\twocolumn

\end{document}